\documentclass[aps,twocolumn,amsmath,amssymb,prl,superscriptaddress,unsortedaddress]{revtex4}

\usepackage{epsf}
\usepackage{graphicx}
\usepackage{color}
\begin{document}

%\title{Isotropic temperature evolution of a phase incoherent $d$-wave pair formation in  Bi2212}
%\title{Observation of quasiparticles for phase incoherent $d$-wave pairing in Bi2212}
\title{Point nodes persisting far beyond $T_{\rm c}$ in Bi2212}

\author{Takeshi Kondo}
\affiliation{ISSP, University of Tokyo, Kashiwa, Chiba 277-8581, Japan}

\author{W.~Malaeb} 
\affiliation{ISSP, University of Tokyo, Kashiwa, Chiba 277-8581, Japan}

\author{Y.~Ishida} 
\affiliation{ISSP, University of Tokyo, Kashiwa, Chiba 277-8581, Japan}

\author{T.~Sasagawa} 
\affiliation{Materials and Structures Laboratory, Tokyo Institute of Technology, Yokohama, Kanagawa 226-8503, Japan}

\author{H.~Sakamoto} 
\affiliation{Department of Crystalline Materials Science, Nagoya University, Nagoya
464-8603, Japan}

\author{Tsunehiro~Takeuchi}
\affiliation{Energy Materials Laboratory, Toyota Technological Institute, Nagoya 468-8511, Japan}

\author{T.~Tohyama} 
\affiliation{Department of Applied Physics, Tokyo University of Science, Tokyo 125-8585, Japan}

\author{S.~Shin} 
\affiliation{ISSP, University of Tokyo, Kashiwa, Chiba 277-8581, Japan}

\date{\today}

\maketitle

{\bf  In contrast to a complex feature of antinodal state, suffering from competing order(s),
the ``pure" pairing gap of cuprates is obtained in the nodal region, which therefore holds the key to the superconducting mechanism. 
One of the biggest question is whether the point nodal state as a hallmark of $d-$wave pairing collapses at $T_{\rm c}$ like the BCS-type  superconductors, or it instead survives above $T_{\rm c}$ turning into the preformed pair state. A difficulty in this issue comes from the small magnitude of the nodal gap, which has been preventing experimentalists from solving it. Here we use a laser ARPES capable of ultrahigh energy-resolution, and detect the point nodes surviving far beyond $T_{\rm c}$ in Bi2212. By tracking the temperature evolution of spectra, we reveal that the superconductivity occurs  when the pair breaking rate is suppressed smaller than the single particle scattering rate on cooling, which governs the value of $T_{\rm c}$ in cuprates. 
}

In cuprates, the energy gap (pseudogap) starts opening at a temperature much higher than $T_{\rm c}$, in some cases above room temperature. 
Many experimental evidences  
\cite{Kondo_competition,Khasanov,Chang_NP,Hashimoto,Eric,Shen_PNAS,Yazdani_PG,He_science,Kondo_MDC} 
point to a competing-order origin, rather than the preformed pair, for the pseudogap observed around the antinode with the maximum energy gap. 
The pseudogap state has been revealed to sharply diminishes toward the antinode and disappear far off the node
by the spectral weight analysis for the ARPES data 
of optimally doped samples \cite{Kondo_MDC}.
This is compatible with the recent results by the scanning tunneling spectroscopy (STM)  and the resonant x-ray scattering,
showing that the density wave Q-vectors are detected only in the Cu-O bond directions \cite{Eric,Yazdani_PG,Damascelli_Bi2201,McElroy_PRL,Yazdani_Qvector}. 
While the whole Fermi surface is eventually dominated by the pseudogap close to the Mott insulating phase \cite{Shen_PNAS,Shi_LSCO,Bi2201_nodeless},
 the ``pure" pairing state seems to be realized around the node at least in the optimally- and over-doped regime \cite{Kondo_MDC,Shen_BCS,Kanigel_NP,Nakayama,Dessau_NP,Dessau_PRB,Yazdani,Fujita}.
  The relevant feature to it would be the spatially homogeneous electronic state seen at the low-bias in the STM spectra, which reflects the nodal momentum region \cite{McElroy_PRL,Yazdani,Eric_Mike}. 
 It strongly contrasts to the highly inhomogeneous spectra at the high-bias associated with the antinodal states. 
 Unveiling the nature of the spectral gap near the node is therefore crucial to elucidate the superconducting mechanism in cuprates.
A difficulty however is the small magnitude of the gap, which has been challenging the experimentalists to investigate.

It has been proposed that the pairing-gap evolution with temperature simply follows the conventional BCS function \cite{Shen_BCS}, and   
 Fermi arcs (disconnected segments of gapless Fermi surface) \cite{Norman_arc} emerge at $T_{\rm c}$ \cite{Shen_BCS,Yazdani,Kanigel_NP,Kanigel_PRL,Nakayama,Shen_PNAS}, marking momentum borders between the superconducting and the competing pseudogap regions \cite{Shen_BCS,Shen_PNAS}. 
However, it seems contradicting the observations of the Nernst and diamagnetic effects above $T_{\rm c}$ \cite{Nernst,Nernst2}, which are viewed as signatures of a phase incoherent superconductivity. 

Recently, a contrasting view was proposed \cite{Dessau_NP,Dessau_PRB,Kondo_MDC}:
its underlying idea is that one should discard the notion of electron quasiparticles, instead pay attention to the density of states, which 
is an effective way of judging the existence of energy gap. 
Accordingly a momentum integration of angle-resolved photoemission spectroscopy (ARPES)
spectra has been performed over a selected part of the momentum space.
This quantity contributed from the nodal region was found to have a gap-like structure even above $T_{\rm c}$ \cite{Dessau_NP,Dessau_PRB}. 
The result seems to be in direct opposition to the above widely accepted view.
Nevertheless, the evidence for single-particle gap with the point nodes surviving above $T_{\rm c}$ is still missing,
and strongly desired to unveil the nature of Fermi arc, which is tiled to the pairing mechanism of cuprates.

The determination of the momentum-resolved gap structure has been also attempted by STM studies through the spectral line-shape analysis \cite{Yazdani}, and by applying the octet model to the interference pattern \cite{Kohsaka,Hanaguri_NP,Fujita}. 
While these STM techniques are very successful, they seem to be limited to the investigation of the antinodal region; 
 the  gap structure close to the node is not determined \cite{Kohsaka,Hanaguri_NP}, 
 or the gapless Fermi arc is obtained even below $T_{\rm c}$ \cite{Yazdani,Fujita}, which has never been reported by the ARPES studies.
It is thus crucial to investigate the nodal region by means of the ARPES, which is specialized for an observation of the momentum space. 

Here we examine the momentum-resolved single-particle spectra of Bi2212 obtained by a laser ARPES \cite{Kiss}. 
 The ultra-high energy resolution and bulk-sensitivity achieved by utilizing a low-energy laser source ($h\nu$=7eV) enabled us to obtain high quality spectra with an extremely sharp line shape. We demonstrate, within the quasiparticle picture, the absence of 
 the gapless Fermi arc at $T_{\rm c}$, and an isotropic temperature evolution of point nodal pairing state 
persistent far above $T_{\rm c}$. 
We find that not only the single-particle scattering rate (${\Gamma _{{\rm{single}}}}$), but also the pair breaking rate (${\Gamma _{{\rm{pair}}}}$) is 
 required to reproduce the ARPES spectra.
Furthermore, the magnitude of $T_{\rm c}$ is determined by the mutual relation between  
${\Gamma _{{\rm{single}}}}$ and ${\Gamma _{{\rm{pair}}}}$ in the form of ${\Gamma _{{\rm{single}}}}(T_{\rm c})$= ${\Gamma _{{\rm{pair}}}}(T_{\rm c})$. 
Importantly, the momentum-integrated spectra of ARPES and STM previously investigated are not capable of separating these two quantities ($\Gamma _{{\rm{single}}}$ and $\Gamma _{{\rm{pair}}}$) (see Supplementary Note 4),  thus the present results provide 
  novel ingredients essential to formulate the pairing mechanism of cuperates.

In Fig.1, we show typical data obtained inside the nodal region where the Fermi arc (bold orange curve in the inset of Fig.1d) was previously claimed to appear at $T_{\rm c}$ \cite{Yazdani,Shen_BCS,Kanigel_NP,Kanigel_PRL,Shen_PNAS}. 
The ARPES intensity map divided by the Fermi function  (see Fig.1a) shows an energy gap and  
an upper branch of the Bogoliubov dispersion at low temperatures, as an indication of the pairing state.
We extract the spectra
at the Fermi momentum ($\bf {k_{\rm F}}$) over a wide range of temperature in Fig.1b, and plot the peak energies ($\varepsilon _{\rm peak}$s) in Fig.1e. 
  In the same panel, we also plot $\varepsilon _{\rm peak}$s of EDCs symmetrized about the Fermi energy ($E_{\rm F}$) to remove the effect of Fermi cut-off \cite{Norman_arc}, and confirm a consistency between the two different methods. 
 Our high quality data clearly exhibit that the 
  gap is open even at $T_{\rm c}$ (=92K) (see green spectra in Fig.1b and 1c), and   
the ${\varepsilon _{\rm peak}}(T)$ disagrees with the BCS gap evolution (blue solid curve in Fig.1e).
 Even if assuming a phase fluctuation slightly above $T_{\rm c}$,  the BCS-type curve (blue dashed curve) still does not fit to the data.
   
To pin down the cause of this anomaly, we examine the momentum variation of ${\varepsilon _{\rm peak}}(T)$ for the optimally and overdoped samples (OP92K and OD72K) in Fig. 2a and Fig. 2b, respectively. 
Surprisingly, the gap does not close at $T_{\rm c}$ regardless of $\bf {k_{\rm F}}$ points. 
 The symmetrized EDCs around the node for OP92K are plotted in Figs. 3b ($T = 10$K) and 3c ($T = T_{\rm c}$). 
We find that the $d$-wave gap with a point node persists at  $T_{\rm c}$ (Fig.3d); 
the Fermi arc is absent. 
While a small uncertainty in gap estimation 
remains in the close vicinity of the node due to the finite spectral width, 
it is negligible compared with the previously 
reported extensive gapless Fermi arc (orange arrows in the inset of Figs. 2a and 2b). 
The absence of Fermi arc is further confirmed in Fermi-function divided band dispersions (Fig. 3a) measured at $T_{\rm c}$ along several momentum cuts (color lines in the inset of Fig. 3d).   
The loss of spectral weight at $E_{\rm F}$ due to the gap opening is seen in all the maps except for at the node (see Supplementary Fig.6 for more details). 
Our high resolution data also show other inconsistencies with the previous expectations \cite{Kanigel_NP,Nakayama,Shen_BCS,Shen_PNAS}. 
First, the length of arc with single spectral-peaks  (${\varepsilon _{\rm peak}}=0$) is not linearly extrapolated to zero at $T$=0 against the nodal liquid behavior (Supplementary Fig.7) \cite{Kanigel_NP,Nakayama}. Secondly, the temperature evolution of such an arc is gradual up to far above $T_{\rm c}$ with no 
 indication of momentum borders separating two distinct states \cite{Shen_BCS,Shen_PNAS}.

For a further examination, we normalize each curve of ${\varepsilon _{\rm peak}}(T)$ to the maximum value at the lowest temperature in the bottom panels of Figs. 2a (OP92K) and 2b (OD72K).
One can confirm that the data are mismatched with the conventional BCS curve (red dashed curves) even in the close vicinity of the node.
More importantly, the ${\varepsilon _{\rm peak}}(T)$ behavior with a steep drop to zero becomes more gradual 
with getting away from the node, and it eventually follows a BCS-type gap function (green curves)
with an onset much higher than $T_{\rm c}$ ($\sim 135$K and $\sim 89$K for OP92K and OD72K, respectively).

To clarify the anomalous feature above $T_{\rm c}$, 
here we investigate another spectral measure, so-called leading-edge of EDC, which is also 
commonly used for a gap estimation.
Figure 4 {\bf a-d} show the non-symmetrized EDCs of OP92K, normalized to the peak intensity of each curve. 
The energy location of spectral leading-edge (${\varepsilon _{\rm LE}}$), at which the spectral intensity becomes half (white allows in {\bf a-d}), is plotted in Fig. 4{\bf e-h} as a function of temperature. 
The nodal values (Fig.4{\bf e}) have a $T$-linear behavior, which is expected for the spectra dominated by the Fermi cut-off effect.
While the nodal ${\varepsilon _{\rm LE}}(T)$ could have a complex behavior \cite{kordyuk} with a poor energy resolution,
such a effect seems not to be observed in our case. 
Hence, even the slightest energy gap could be detected as the deviation of ${\varepsilon _{\rm LE}}(T)$ from the $T$-linear behavior.
Such a deviation  is indeed observed for all the ${\bf k}$ points off the node (Fig.4{\bf f-h}).
It is more clearly demonstrated in the bottom panels of Fig.4{\bf f-h}, where
the difference of ${\varepsilon _{\rm LE}}(T)$ from the $T$-linear behavior is extracted. 
An astonishing result we found is that the onset temperatures are almost the same ($\sim135$K) regardless of the directions ($\phi$s), which validates that the point nodal state persists up to $\sim 135$K. 
 
 Importantly, the gap opening temperature is coincident with the onset of the BCS-type gap function obtained 
with the symmetrized EDCs far off the node (a green curve in Fig.2a).
A plausible explanation for it is that the spectral peak energy underestimates the ``real" energy gap ($\varepsilon _{\rm peak}<\Delta$) \cite{Varma,Chubukov,Kondo_MDC}, and energy gaps ($\Delta$s) comparable to or smaller than the peak width cannot be detected by tracing the peak positions (see a simulation in the Supplementary Fig.8). This situation could occur at high temperatures, and it gets more serious toward the node with smaller $\Delta$. 
We actually detect the signature of such a small gap 
in the off-nodal spectra with a single peak above $T_{\rm c}$ (Supplementary Fig. 10);
the spectral width becomes smaller with increasing temperature up to $\sim135$K,
which contrasts to the monotonic broadening seen in the nodal direction. 
The characteristic momentum variation of ${\varepsilon _{\rm peak}}(T)$ in Figs. 2a and 2b, therefore, could be a natural consequence of $\Delta (T)$ having the same onset temperature at $\sim 135$K and  $\sim 89$K, respectively, for all directions ($\phi$s).

We find below that a model spectral function, $\pi A({\bf {k_{\rm F}}},\omega ) = \Sigma ''/[{(\omega  - \Sigma ')^2} + {{\Sigma ''}^2}]$, with such a BCS-type $\Delta(T,\phi)$ indeed reproduces the ARPES spectra, whereas the traditionally used assumption of $\Delta (T,\phi) \equiv {\varepsilon _{\rm{peak}}}(T,\phi)$ is invalid. 
The self-energy ($\Sigma  = \Sigma ' + i\Sigma ''$) we use has a minimal representation with 
 two different scattering rates, $\Gamma_{\rm 0}$ and $\Gamma_{\rm 1}$ \cite{Norman_PRB,Chubukov},
\begin{equation} \label{self}
\Sigma ({\bf {k_{\rm F}}},\omega ) =  - i{\Gamma _{{\rm{1}}}} + {\Delta ^2}/[\omega  + i{\Gamma _{{\rm{0}}}}].
\end{equation}
Here $\Gamma_{\rm 1}$ is a single-particle scattering rate, and 
it causes the broadening of a peak width.  On the other hand, $\Gamma_{\rm 0}$ fills the spectral weight around $E_{\rm F}$, 
and should be viewed as the inverse pair lifetime (or pair breaking rate). For clarity, we label the former  $\Gamma_{\rm single}$ and the latter  $\Gamma_{\rm pair}$ in the rest of this paper. 
We emphasize that the intensity at $E_{\rm F}$ in a gapped spectrum becomes non-zero only when ${\Gamma _{{\rm{pair}}}} \ne 0$ as simulated in Supplementary Fig. 8a.
Our spectra measured at the low temperatures ($T \ll {T_{\rm c}}$) have a negligible intensity at $E_{\rm F}$, which 
 ensures that our data are almost free from  impurity-causing pair breaking effect.
At elevated temperatures, we observe a remarkable gap filling (see Fig.1d and Supplementary Figs. 2 and 5). 
Significantly, it actually begins from deep below $T_{\rm c}$, which is not expected in a conventional BCS superconductor. 
Since the data were measured at the extremely high energy resolution ($\Delta\varepsilon$ = 1.4 meV), we can rule out the possibility, assumed before with setting ${\Gamma _{{\rm{pair}}}} \equiv 0$ \cite{Shen_BCS,Shen_PNAS},
that the filling is caused by a spectral broadening due to the experimental energy resolution.
The intensity at $E_{\rm F}$ should instead be a signature of intrinsic pair breaking, hence it must be taken into account for the gap estimation. 
In passing, we note that the $\Gamma_{\rm single}$ and $\Gamma_{\rm pair}$ both  equally increase the intensity around 
$E_{\rm F}$ of the momentum-integrated spectrum previously studied by ARPES \cite{Dessau_NP,Dessau_PRB,Kondo_MDC} and STM \cite{Yazdani,Yazdani_gamma} (see Supplementary Fig. 9), which is therefore incapable of disentangling these two different scattering rates. 

Following this consideration, we set  ${\Gamma _{{\rm{single}}}}$ and   ${\Gamma _{{\rm{pair}}}}$  to be independent free parameters in Eq.(1). 
First,  we performed a spectral fitting to our ARPES data, assuming $\Delta (T) \equiv {\varepsilon _{{\text{peak}}}}(T)$, which is a traditional way of gap estimation (Supplementary Fig. 11a and Fig. 12a). 
The obtained parameter of ${\Gamma _{{\rm{single}}}}(T)$ (middle panel of Supplementary Fig. 11a and Fig. 12a) is strongly deviated from a monotonic decrease on cooling, having an unrealistic upturn around the temperature at which  the ${\varepsilon _{\rm{peak}}}$ becomes zero. 
As already discussed above, this anomaly is expected when the spectrum with a single peak (${\varepsilon _{\rm{peak}}}=0$) has an energy gap ($\Delta \ne 0$), thus the spectral width overestimates the scattering rate. 

We find that this circumstance is corrected by applying 
a BCS-type gap function with an onset at 135K and 89K for OP92K and OD72K (green curves in the bottom panels of Fig.2a and Fig.2b), respectively,
regardless of the Fermi angle $\phi$.
In the Supplementary Fig. 14 and Fig. 15, 
we fit  Eq.(1) with such a gap function $\Delta$ to our ARPES data 
near the node measured over a wide temperature range.
As an example, the result at $\phi=13.5^\circ$ for OP92K
 is shown in Fig.5c.
The fitting curves (red curves) almost perfectly reproduce the data (black curves).
The obtained ${\Gamma _{{\rm{single}}}}(T)$s (Fig. 5a and 5b) in the gapped region agree with 
$\Gamma_\mathrm{single} (T)$ at the node, which can be determined simply from the spectral width.
Similarly, the ${\Gamma _{{\rm{pair}}}}(T)$ curves  are also almost identical  for all the $\phi$s.
The consistency in our results
pointing to the isotropic scattering mechanism
validates our model spectral function characterized by the BCS-type $\Delta(T,\phi)$. 
The famous ``hot spots", at which the scattering rate is abruptly enhanced, should be situated at much higher $\phi$s (Supplementary Fig. 13).
The applied onset temperatures are almost the same as 
those of Nernst and diamagnetic effects \cite{Nernst,Nernst2}, that are viewed as signatures of phase-incoherent superconductivity. 
The comparable temperatures are also obtained by the specific heat measurements \cite{Tallon} and the other spectroscopic techniques \cite{Kondo_pair,Yazdani_pair}.
Therefore, we assign $\sim 135$K and $\sim 89$K to be the onset temparature of pair formation ($T_{\rm pair}$) of OP92K and OD72K, respectively. 
This is further supported by the  signature of pairing  seen in 
the behavior of ${\Gamma _{{\rm{single}}}}(T)$ (Fig.5a and 5b, and Supplementary Fig. 16);  
the decrease of its value upon cooling is accelerated across $T_{\rm pair}$, showing a deviation from the linear behavior \cite{barivsic2013universal}. 
The different experimental techniques could have different sensitivities to the superconducting fluctuation above $T_{\rm c}$, and actually the terahertz spectroscopy estimates a slightly lower temperature scale  (10-15K above $T_{\rm c}$) \cite{THz1,THz2}.  Nonetheless, we stress that the view that the point-nodal pairing survives above $T_{\rm c}$ is compatible in these observations.
The doping variation of $T_{\rm pair}$/$T_{\rm c}$  (1.47 and 1.24 for OP92K and OD72K, respectively)  obtained  in our studies
is consistent with the phase-fluctuating superconductivity, which merges to the superconducting dome with heavily overdoping \cite{Emery}. The competing ordered-phase is, in contrast, claimed to terminate at zero temperature inside the superconducting dome \cite{shekhter2013bounding}, thus disagrees with the present gapped states observed above $T_{\rm c}$ even in the overdoped sample. 
 
The relationship between ${\Gamma _{{\rm{pair}}}}$ and ${\Gamma _{{\rm{single}}}}$  should provide rich information relevant for the pairing mechanism.
Intriguingly, the superconductivity occurs when the magnitude of ${\Gamma _{{\rm {pair}}}}$ is reduced smaller than that of  ${\Gamma _{{\rm {single}}}}$; 
the $T_{\rm c}$  is coincident with the temperature at which ${\Gamma _{{\rm{single}}}}(T)$ and  ${\Gamma _{{\rm{pair}}}}(T)$ crosses (magenta circles in Fig. 5a and 5b), which provides a simple formula of ${\Gamma _{{\rm{single}}}}(T_{\rm c})$= ${\Gamma _{{\rm{pair}}}}(T_{\rm c})$.
The magnitude of $T_{\rm pair}$  is reported to be comparable (120K$\sim$150K)  among different cuprate families with significantly different $T_{\rm c}$s \cite{Kondo_pair}.
The ${\Gamma _{{\rm{single}}}}(T)$ also seems to be less sensitive to the different compounds \cite{kondo_kink,Zhou}. Therefore, 
the pair breaking effect, which controls the  fulfillment of ${\Gamma _{{\rm {pair}}}} < {\Gamma _{{\rm {single}}}}$, 
 is predicted to be a critical factor determining the $T_{\rm c}$ value of cuprates.
 Notably, a remarkable difference in filling behaviors of the spectral gap is indeed observed between  Bi2212 and Bi$_2$Sr$_2$CuO$_{6+\delta}$ with about three times different $T_{\rm c}$s \cite{Kondo_pair,Kondo_MDC}. 
 
We summarize our conclusion in Fig.6 by drawing a schematic temperature evolution of the pairing gap. 
As demonstrated elsewhere \cite{Kondo_MDC}, the pseudogap competing with the pairing disappears far off the node (blue area).
In this article, we have investigated the nodal region of optimally and overdosed samples. 
We revealed that the point-node state persists up to far above $T_{\rm c}$, against the previous expectation \cite{Shen_BCS,Nakayama,Kanigel_NP,Kanigel_PRL,Yazdani,Fujita}, following the BCS-type gap function with the onset at $T_{\rm pair}$ ($\sim 1.5T_{\rm c}$ in the optimal doping). The ARPRS spectra are reproduced by the spectral function with a minimal model, which has a single energy gap all the way up to the gap closing temperature. It is consistent with the  expected ``pure" pairing state with no contamination by the pseudogap around the node at least in the  optimally and overdosed regions. 
While the gap evolution across $T_{\rm pair}$ might be reminiscent of a phase-transition phenomenon, it is against the specific heat data, showing a gradual variation with temperature above $T_{\rm c}$ \cite{Tallon}.
The crossover-like behaviors in cuprates  should come from the significant pair breaking effect, which dramatically enhances across $T_{\rm c}$.  
 To fully understand the present results, insight of the spacially inhomogeneous state \cite{Pan}
would be essential. However, only that cannot explain our data, since 
the local density of states itself has the behavior of gap filling at $E_{\rm F}$ with temperature \cite{Yazdani_gamma}. 
The competing nature of pseudogap state evolving around the antinode \cite{Kondo_competition,Chang_NP} is a plausible source for 
the unique scattering mechanism, which strongly suppress $T_{\rm c}$ from $T_{\rm pair}$. 
To evaluate this speculation, however, the more detailed theoretical inputs are required.

\textbf{Methods}

Optimally doped Bi$_2$Sr$_2$CaCu$_2$O$_{8+\delta}$ (OP92K) and overdoped (Bi,Pb)$_2$Sr$_2$CaCu$_2$O$_{8+\delta}$ (OD72K)
single crystals with $T_{\rm c}$=92K and 72K, respectively,
were grown by the conventional floating-zone (FZ) technique. 
A sharp superconducting transition width of  $\sim$1 K (OP92K) and $\sim$3 K (OD72K) were confirmed (see Supplementary Fig. 1).  

ARPES data was accumulated using a laboratory-based system consisting of a Scienta R4000 electron analyzer and  a 6.994 eV laser. The overall energy resolution in the ARPES experiment was set to 1.4 meV for all the measurements.  In order to accomplish the temperature scan of spectra at a high precision, 
we applied a technique of the local sample heating, 
which thermally isolates the sample holder with a heat switch from the lest of the system at elevated temperatures. 
It minimizes the degassing, allowing us to keep the chamber pressure better than $2 \times {10^{ - 11}}$ torr
during the entire temperature sweeping; no sample aging was confirmed (Supplementary Fig. 4). 
This method also prevents the thermal expansion of sample manipulator, and it enables us to
take data in fine temperature steps with automated measurement of temperature scan from precisely the same spot on the crystal surface, which was essential to achieve the aim of the present study.
 
 {\bf  Acknowledgements}\\
  We thank M. Imada, S. Sakai and T. Misawa for useful discussions. This work is supported by JSPS (KAKENHI Grants No. 24740218 and No. 25220707, and FIRST Program).

%\bibliography{Bi2212_arc}

%\clearpage

%%%%%%%%%%%%%%%%%%%%%%
\begin{figure*}
\includegraphics[width=6.5in]{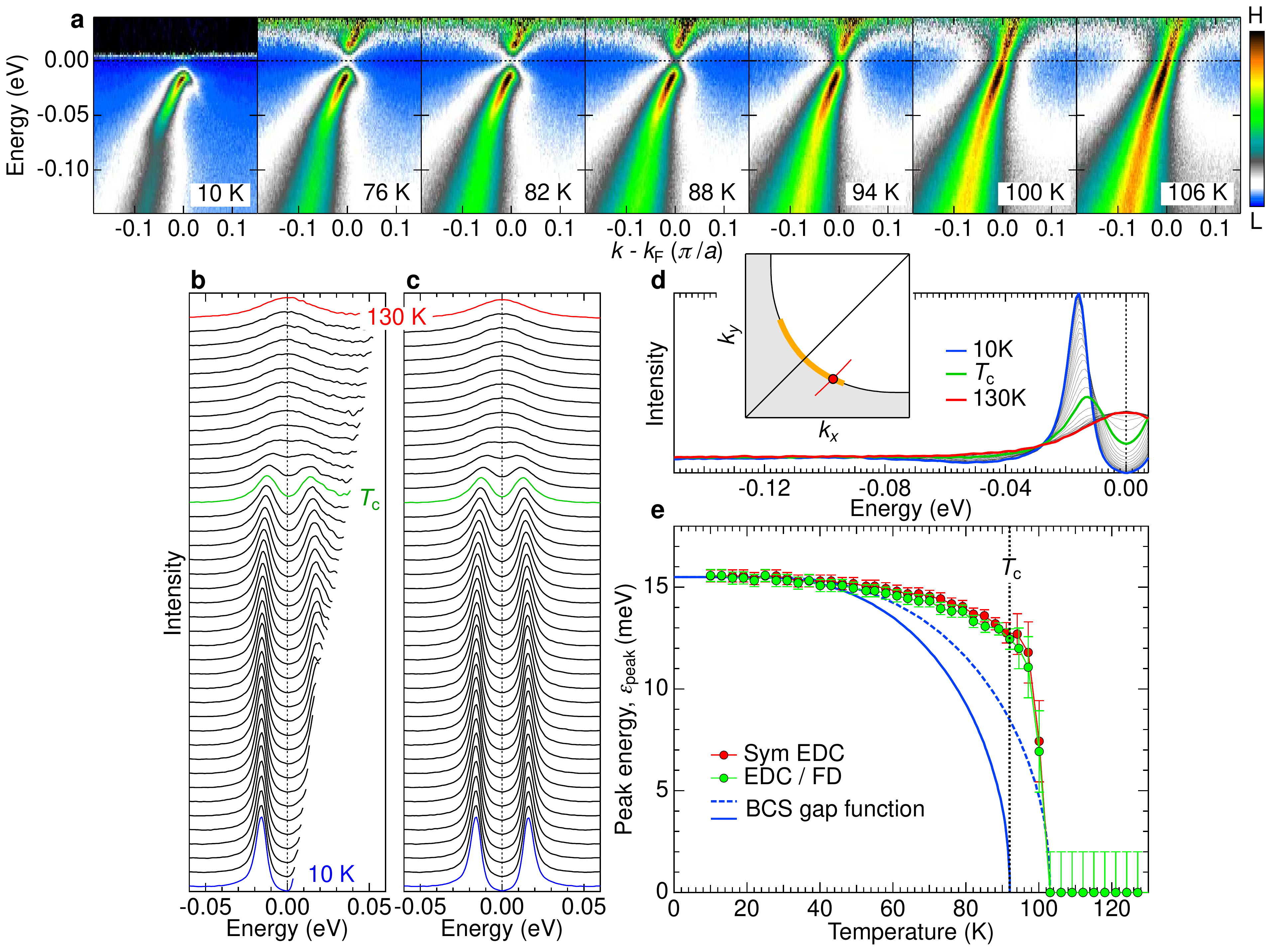}
\caption{ {\bf Temperature evolution of ARPES spectra in the nodal region.}
 {\bf a}, Dispersion maps at several temperatures measured along a momentum cut 
 close to the node in OP92K 
 (a red line in the inset of {\bf d}). 
Each map is divided by the Fermi function at the measured temperature. 
{\bf b}, Temperate evolution of EDCs at $\bf {k_{\rm F}}$ (a circle in the inset of {\bf d}) from deep below  (10K) to much higher than $T_{\rm c}$ (130K). 
Each spectrum is divided by the Fermi function at the measured temperature. 
{\bf c}, The same data as in {\bf b}, but symmetrized about $E_{\rm F}$. 
{\bf d}, The same data as in {\bf c} plotted without an offset. The inset represents the Fermi surface.
The bold orange line indicates the momentum region where the Fermi arc was previously claimed  to emerge at $T_{\rm c}$.  
{\bf e}, Peak energies of spectra in {\bf b} and {\bf c} plotted as a function of temperature, ${\varepsilon _{\rm peak}}$. 
The  solid and dashed blue curves show the BCS gap function with an onset at $T_{\rm c}$ (92K) and slightly above $T_{\rm c}$, respectively. 
Error bars in {\bf e} represent uncertainty in estimating the spectral peak positions. 
} 
\label{fig1}
\end{figure*}
%%%%%%%%%%%%%%%%%%%%%%

%\clearpage

%%%%%%%%%%%%%%%%%%%%%%
\begin{figure*}
\includegraphics[width=6.5in]{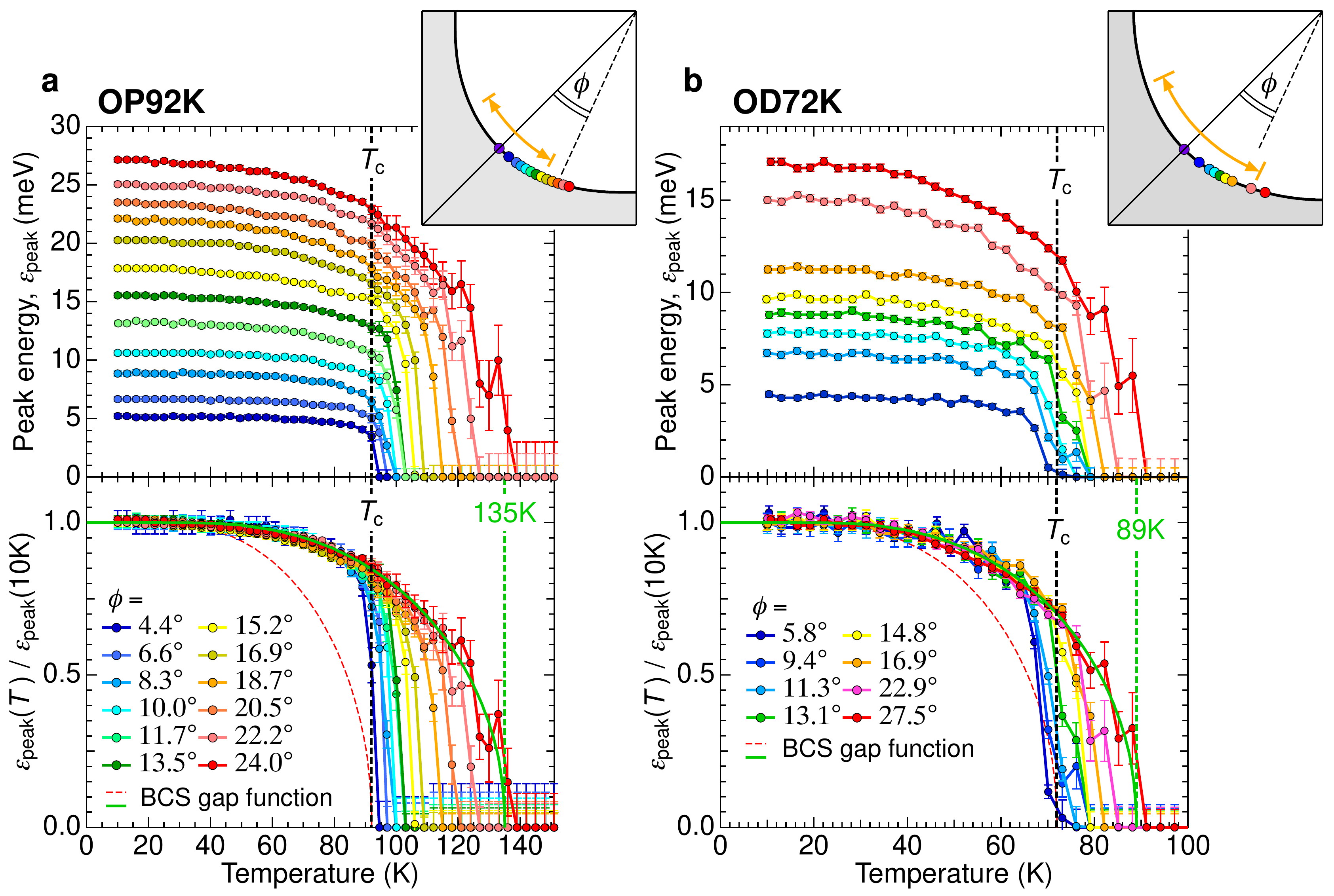}
\caption{{\bf Disagreement between the ARPES results and the conventional BCS gap function.}
{\bf a,b}, Temperature dependence of spectral peak energy, ${\varepsilon _{\rm peak}}(T)$, at various $\bf {k_{\rm F}}$ points (color circles in the insets) for OP92K and OD72K, respectively. In the bottom of each panel, the same curves of ${\varepsilon _{\rm peak}}(T)$ are normalized to the maximum value at the lowest temperature.  
A red dashed curve and a green solid curve are the BCS gap function with the onset at $T_{\rm c}$ and $T_{\rm pair}$, respectively ($T_{\rm pair}=$135K for OP92K and  $T_{\rm pair}=$89K for OD89K). 
The inset shows the Fermi surface with measured $\bf {k_{\rm F}}$ points (color circles).  
The bold orange line indicates the momentum region where the Fermi arc was previously claimed  to emerge at $T_{\rm c}$. 
Error bars in {\bf a} and {\bf b} represent uncertainty in estimating the spectral peak positions. }
\label{fig1}
\end{figure*}
%%%%%%%%%%%%%%%%%%%%%%

%\clearpage

%%%%%%%%%%%%%%%%%%%%%%
\begin{figure*}
\includegraphics[width=6.5in]{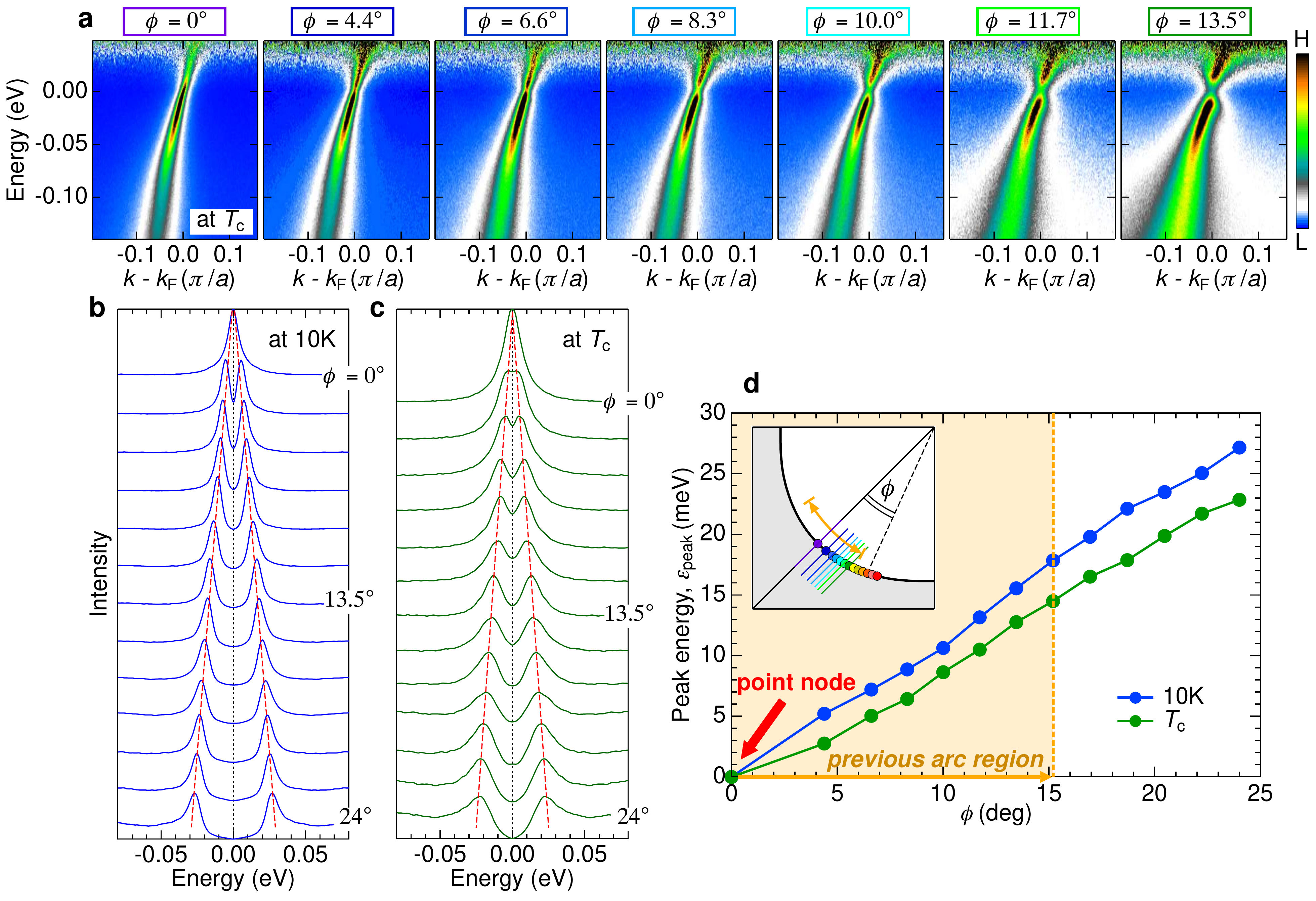}
\caption{{\bf Momentum variation of spectra showing the absence of  Fermi arc at $T_{\rm c}$.}
{\bf a}, Dispersion maps at $T_{\rm c}$ along several momentum cuts  (color lines in the inset  of {\bf d}) measured for OP92K. 
Each map is divided by the Fermi function at the measured temperature.  
The described $\phi$s are the directions of measured $\bf {k_{\rm F}}$ points ($\phi$ is defined in the inset  of {\bf d}). 
{\bf b},{\bf c},  Symmetrized EDCs at $\bf {k_{\rm F}}$ 
over a wide range of angle $\phi$ (color circles in the inset  of {\bf d}) at 10K and $T_{\rm c}$ (=92K), respectively.
The red dotted lines are guide to the eyes for the gap evolution.
{\bf d}, Fermi angle $\phi$ dependence of 
peak energies of spectra in {\bf b} (${\varepsilon _{\rm peak}}$(10K)) and {\bf c} (${\varepsilon _{\rm peak}}(T_{\rm c})$). 
The orange arrows indicate the momentum region where the Fermi arc was previously claimed  to emerge at $T_{\rm c}$.} 
\label{fig1}
\end{figure*}
%%%%%%%%%%%%%%%%%%%%%%

%\clearpage

%%%%%%%%%%%%%%%%%%%%%%
\begin{figure*}
\includegraphics[width=6.5in]{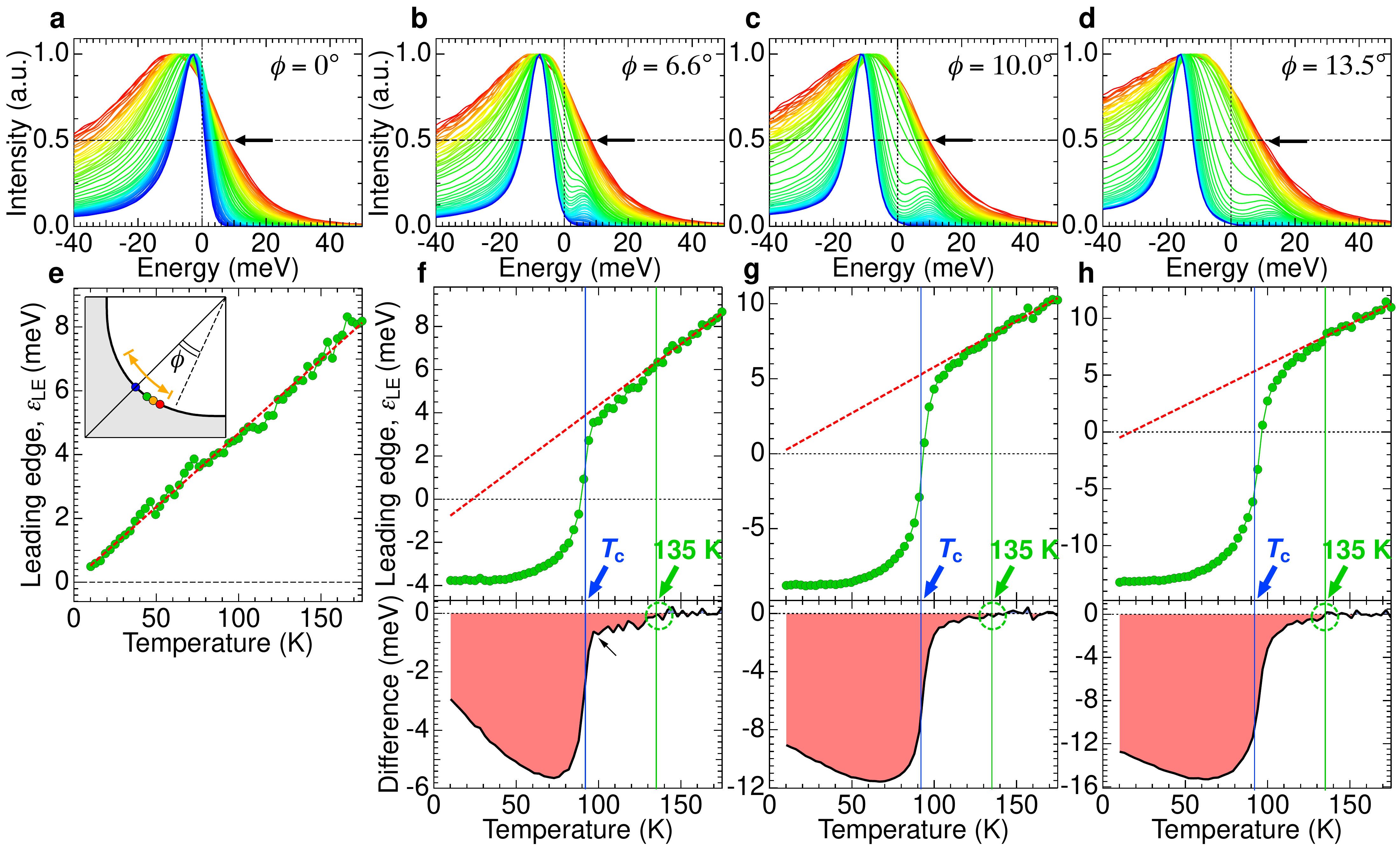}
\caption{{\bf Evidence for the point nodal gap surviving above  $T_{\rm c}$  revealed by means of the energy shift of spectral leading-edge.}
{\bf a}, The temperature evolution of ARPES spectra (EDCs) at the gap node for OP92K, normalized to the peak intensity of each curve.
{\bf b-d} The same spectra as {\bf a}, but measured off the node (color circles in the inset of {\bf e}). 
{\bf e-h} Temperature dependence of the spectral leading-edge, $\varepsilon_{\rm LE}(T)$, defined as energies where the spectral peaks become half in intensity (marked by white arrows in {\bf a-d}).
The bottom panels of {\bf f-h} plot the difference of $\varepsilon_{\rm LE}(T)$  from the $T$-linear behaviors (red dashed lines), which are fit to  $\varepsilon_{\rm LE}(T)$ at high temperatures. 
The onset temperaturess of the deviation from $T$-linear  behavior are around 135K for all the $\bf {k_{\rm F}}$ points off the node (yellow dashed circles).
A kink is seen in the difference curve for {\bf f} (small black arrow) because the thermally populated Bogoliubov peaks in the unoccupied side significantly affect the shapes of leading-edge 
specially at $\bf {k_{\rm F}}$s close to the node with small gaps. 
The inset of {\bf e} shows the Fermi surface with the examined $\bf {k_{\rm F}}$ points.
The bold orange arrow indicates the momentum region where the Fermi arc was previously claimed  to emerge at $T_{\rm c}$. 
} 
\label{fig1}
\end{figure*}
%%%%%%%%%%%%%%%%%%%%%%

%\clearpage

%%%%%%%%%%%%%%%%%%%%%%
\begin{figure*}
\includegraphics[width=5.5in]{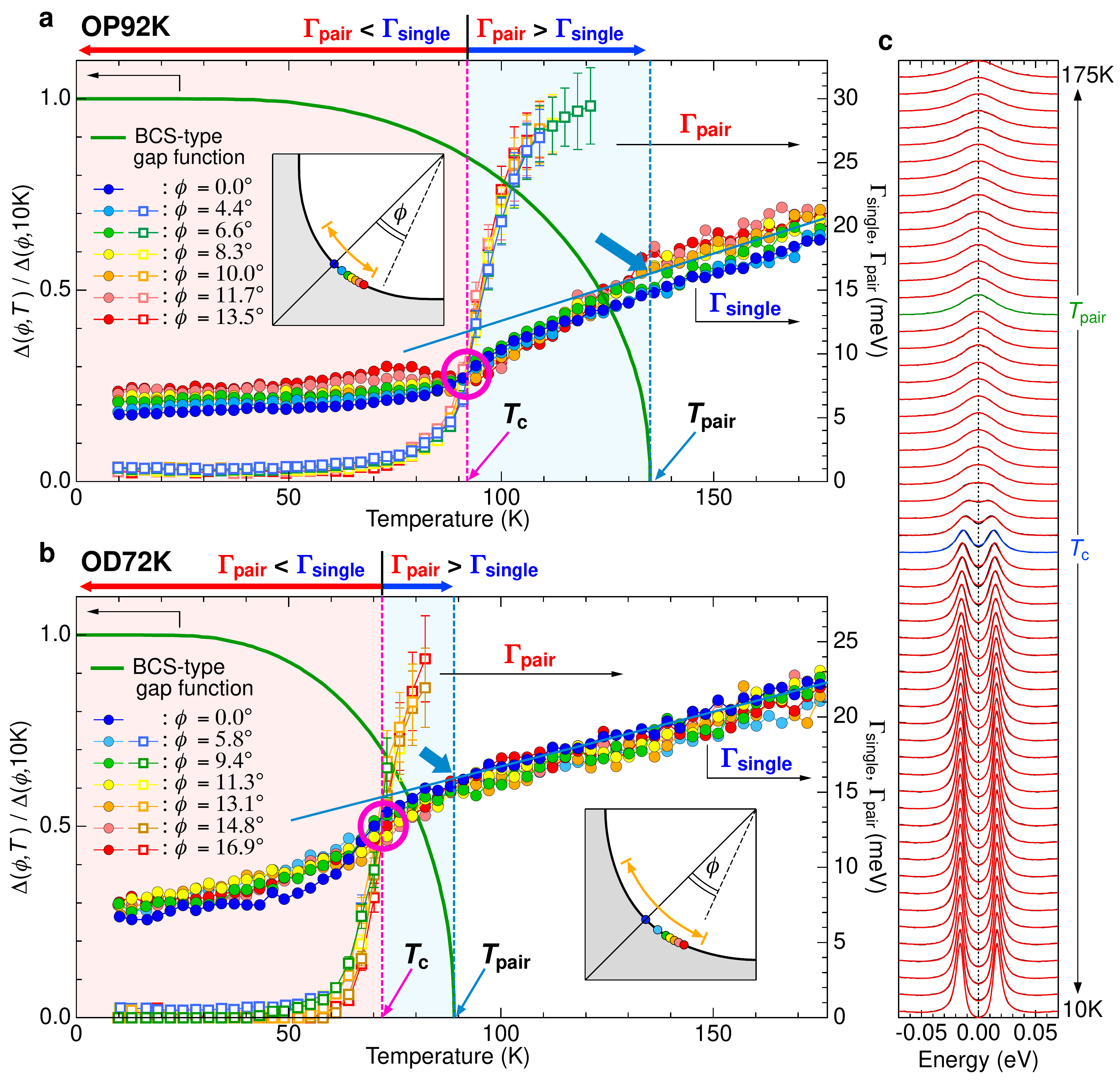}
\caption{ {\bf Extracted parameters of the minimal model spectral function required to reproduce the ARPES spectra.}
{\bf a}, The BCS-type gap function used for the fitting to the OP92K data (a green curve), and the obtained 
single particle scattering rate ($\Gamma_{\rm single}$) and the pair breaking rate ($\Gamma_{\rm pair}$) in Eq.(1). 
{\bf b}, The fitting results same as in {\bf a}, but for the OD72K data. 
The values of $\Gamma_{\rm pair}$ at high temperatures are not plotted, since the spectral shape is insensitive to the $\Gamma_{\rm pair}$  when $\Delta$ is small or zero, and thus it is impossible to  determine the value.
A small hump seen in the $\Gamma_{\rm pair}$ around 75K for $\phi  = 11.7^\circ$ and $13.5^\circ$ in {\bf a} comes from a slight difficulty of fitting to the spectra with a peak-dip-hump shape due to the mode coupling, which appears below $T_{\rm c}$ and gets pronounced with approaching the antinode.  Magenta circle marks the crossing point of $\Gamma_{\rm single}(T)$ and $\Gamma_{\rm pair}(T)$. 
Large blue arrow indicates the temperature, at which the  $\Gamma_{\rm single}(T)$ deviates from a $T$-linear behavior upon cooling.
{\bf c}, ARPES spectra (black curves) and fitting results (red curves) providing the parameters for $\phi  = 13.5^\circ$ in {\bf a}. 
 Error bars in {\bf a} and {\bf b} represent uncertainty in fitting the model spectral function to the ARPES spectra. } 
\label{fig1}
\end{figure*}
%%%%%%%%%%%%%%%%%%%%%%

%\clearpage

%%%%%%%%%%%%%%%%%%%%%%
\begin{figure*}
\includegraphics[width=5.5in]{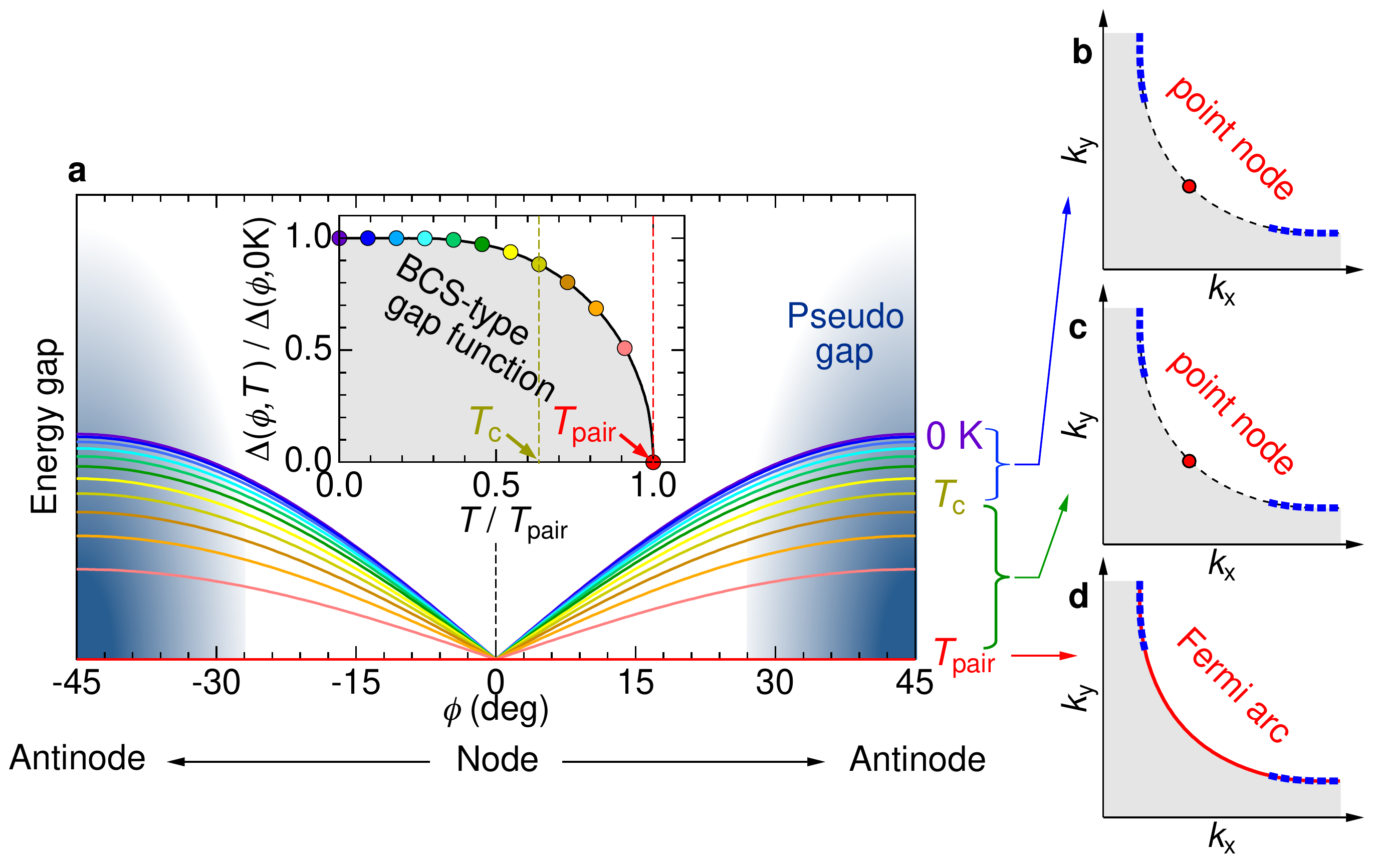}
\caption{ {\bf  Schematic pairing-gap evolution based on our ARPES results. }
{\bf a}, Temperature dependence of the point nodal $d$-wave pairing gap with BCS-type temperature dependence (inset curve) regardless of directions ($\phi$s) along the Fermi surface. Temperatures for each curve are indicated in the inset with colored circles. The gapped Fermi surface with a point node   below $T_{\rm c}$ ({\bf b})
persists   beyond $T_{\rm c}$  ({\bf c}) up to the temperature of pair formation ($T_{\rm pair}$). 
{\bf d},  Emergence of the gapless Fermi arc centered at the node due to the pseudogap  evolution around the antinode \cite{Kondo_MDC,Kohsaka}; 
while the antinodal region is not observable at the low photon energies like 7eV, 
the studies with higher energy photons  have demonstrated that the competing pseudogap state emerges  at
$\left| \phi  \right| > 25^\circ$ in the optimal doping \cite{Kondo_MDC} .
 } 
\label{fig1}
\end{figure*}
%%%%%%%%%%%%%%%%%%%%%%

\end{document}

\clearpage
 \newpage
 
 \title{Supplementary information for\\
Point nodes persisting far beyond $T_c$ in Bi2212}

\author{Takeshi Kondo}
\affiliation{ISSP, University of Tokyo, Kashiwa, Chiba 277-8581, Japan}

\author{W.~Malaeb} 
\affiliation{ISSP, University of Tokyo, Kashiwa, Chiba 277-8581, Japan}

\author{Y.~Ishida} 
\affiliation{ISSP, University of Tokyo, Kashiwa, Chiba 277-8581, Japan}

\author{T.~Sasagawa} 
\affiliation{Materials and Structures Laboratory, Tokyo Institute of Technology, Yokohama, Kanagawa 226-8503, Japan}

\author{H.~Sakamoto} 
\affiliation{Department of Crystalline Materials Science, Nagoya University, Nagoya
464-8603, Japan}

\author{Tsunehiro~Takeuchi}
\affiliation{Energy Materials Laboratory, Tokyota Technological Institute, Nagoya 468-8511, Japan}

\author{T.~Tohyama} 
\affiliation{Department of Applied Physics, Tokyo University of Science, Tokyo 125-8585, Japan}

\author{S.~Shin} 
\affiliation{ISSP, University of Tokyo, Kashiwa, Chiba 277-8581, Japan}

\date{\today}

\maketitle

{\bf Supplementary Note 1: Samples and Experimental method}\\

Optimally doped Bi$_2$Sr$_2$CaCu$_2$O$_{8+\delta}$ (OP92K) and overdoped (Bi,Pb)$_2$Sr$_2$CaCu$_2$O$_{8+\delta}$ (OD72K)
single crystals with $T_{\rm c}$=92K and 72K, respectively,
were grown by the conventional floating-zone (FZ) technique. 
Magnetic susceptibilities for these single crystals are shown in Supplementary Figure \ref{SQUID}. 
Sharp superconducting transitions with $\sim$1 K (OP92K) and $\sim$3 K (OD72K), indicative of a high quality, are confirmed.  
ARPES data were accumulated using a laboratory-based system consisting of a Scienta R4000 electron analyzer and  a 6.994 eV laser (the 6th harmonic of Nd:YVO4 quasi-continuous wave with a repetition rate of 240 MHz). The overall energy resolution in the ARPES experiment was set to 1.4 meV for all the measurements.
   
In order to accomplish the temperature scan of spectra at a high precision, 
we made a special sample stage;
we use a heat switch to thermally isolate the sample holder with a heater and a temperature sensor (totally $20 \times 30 \times 10$ mm size)  
from the rest of the system when sweeping the sample temperature.
With this technique, the samples are efficiently heated up  from 10K up to 300K 
while keeping the temperature of manipulator rod other than the sample stage lower than 35K. 
It minimizes the degassing  
and keeps the  pressure in the measurement chamber better than $2 \times {10^{ - 11}}$ torr
during the entire temperature sweeping. 
The method of local heating also prevents the thermal expansion of the $long$ manipulator rod, and thus the sample position becomes unchanged with temperature. 
It enabled us to
take data in fine temperature steps with the automated measurement of temperature scan from precisely the same spot on the crystal surface, which were essential to achieve the aim of this study.

In Supplementary Figs. \ref{WithoutOffsetOP92K} and \ref{WithOffsetOP92K}, we show the ARPES spectra of OP92K obtained at various $\bf {k_{\rm F}}$ points with sweeping the temperature, with and without an offset, respectively. These are the original data 
used for the analysis in the main paper. 
The ARPES technique is sensitive to the condition of sample surface, and the biggest challenge in it is to 
prevent the aging of sample surface with time. 
In Supplementary Figure \ref{AgingCheck}, we carefully check the stability of sample surface in our experiment 
by comparing two sets of spectra measured before and after the temperature scan for all the momentum cuts we measured. 
While it took about 12 hours to complete each temperature sweeping, 
 almost perfect reproducibility of spectral line shapes is confirmed for all the data.
This ensures that our data measured with the new heating technique 
are reliable and suitable for conducing the detailed analysis presented in the paper. 
  Supplementary Figure \ref{WithoutOffsetOD72K} shows the data set similar to those in Supplementary Figure \ref{WithOffsetOP92K}, but measured for OD72K.
\\\\

 {\bf Supplementary Note 2: Further Evidence for absence of the Fermi arc at $T_{\rm c}$}\\

 In the main paper, we demonstrate that the $d$-wave gap with a point node persists at $T_{\rm c}$ by plotting the symmetrized EDCs at various $\bf {k_{\rm F}}$ points around the node (see Fig. 3d). Here we present a further evidence that the conclusion is robust and it is not sensitive to the selection of $\bf {k_{\rm F}}$ value. 
 Supplementary Figure \ref{Pointnode}a shows  the Fermi-function divided band dispersions (same as Fig.3a in the main paper) measured at $T_{\rm c}$ along several momentum cuts (color lines in Supplementary Figure \ref{Pointnode}c). The corresponding EDCs are plotted in Supplementary Figure \ref{Pointnode}b. 
Along the diagonal cut (the most left panel), the spectra with one peak crosse $E_F$ due to the gap node. 
In contrast, two peak structure (marked with arrows), signifying a gap existence, is observed at $\bf {k_{\rm F}}$ (painted with red) for the other directions. This is a direct evidence that the $d$-wave gap with a point node persists, thus the Fermi arc is absent at $T_{\rm c}$.
\\\\

 {\bf Supplementary Note 3: Temperature evolution of the ``artificial" Fermi arc}\\

The Fermi arc \cite{Norman_arc} is one of the most interested features in cuprates. 
Previous studies have proposed that the gapless Fermi arc abruptly emerges at $T_{\rm c}$, and 
expands linearly with a further increase of temperature \cite{Shen_BCS,Kanigel_NP,Nakayama}.
The arc length was claimed to be extrapolated to zero at $T=0$ \cite{Kanigel_NP,Nakayama}. 
The so-called ``nodal liquid" behavior has been having a huge influence on the condensed matter community. 

In the main paper, we argue that the tracking of  peak positions of the symmetrized EDCs ($\varepsilon _{\rm peak}$s) underestimates the gap closing temperature, thus provides an ``artificial" Fermi arc. 
Here we demonstrate that even such an artificial arc behaves quite differently from the previous expectation,
 when high resolution data by a laser ARPES are examined. 
In Supplementary Figure \ref{ArcLength}a and  \ref{ArcLength}a,  we estimate the length of arc with $\varepsilon _{\rm peak}$=0 as a function of temperature for OP92K and OD72K, respectively.
In the same panels, the nodal liquid behaviors are superimposed.
Our results are significantly off these lines. 
As discussed in the main paper, the spectral line-shape is determined not only by  the energy gap ($\Delta$), but also by the mutual relation between the single-particle scattering rate ($\Gamma_{\rm single}$) and the pair breaking rate ($\Gamma_{\rm pair}$). Therefore, it is rather reasonable that the growth of arc with $\varepsilon _{\rm peak}$=0 becomes arbitrary.
\\\\

  {\bf Supplementary Note 4: Effect of scattering rates, $\Gamma_{\rm single}$ and $\Gamma_{\rm pair}$, on the spectral line shape}\\

In the analysis, we used the spectral function with a phenomenological self-energy (ref. \cite{Norman_PRB}) given by  
\renewcommand {\theequation}{S\arabic{equation}}
\begin{eqnarray} \label{self_Norman}
A({\bf k},\omega ) = \frac{1}{\pi } \cdot \frac{{\Sigma ''({\bf k},\omega )}}{{{{\left[ {\omega  - \varepsilon ({\bf k}) - \Sigma '({\bf k},\omega )} \right]}^2} + \Sigma ''{{({\bf k},\omega )}^2}}},\nonumber \\ \nonumber \\
\Sigma ({\bf k},\omega ) =  - i{\Gamma _{{\rm{single}}}} + {\Delta ^2}/[\omega  + \varepsilon ({\bf k}) + i{\Gamma _{{\rm{pair}}}}].
\end{eqnarray}  
Here $\Delta$ is the energy gap, and  $\varepsilon (k)$ the energy dispersion with $\varepsilon (\bf {k_{\rm F}})=$0. $\Gamma_{\rm single}$ and $\Gamma_{\rm pair}$ are the single particle scattering rate and the pair breaking rate, respectively.
In Supplementary Figure \ref{Simulation}, we examine the effect of these scattering rates 
on the spectral line shape at $\bf {k_{\rm F}}$ with fixing $\Delta$  to 10 meV.
To make it realistic, 
all the curves plotted are convoluted with a Gaussian that has the width of  the experimental energy resolution ($\Delta \varepsilon  = 1.4$meV). Nevertheless, note that the $\Delta \varepsilon$ value  is so small  that the difference in  shape from original curves is negligible.  
Figure \ref{Simulation}a shows the $\Gamma_{\rm single}$ dependence of the spectral function when  the $\Gamma_{\rm pair}$ is fixed to zero.
One finds that the intensity at $E_{\rm F}$ is always zero regardless of the $\Gamma_{\rm single}$ value, whereas the spectral width gets broadened with increasing $\Gamma_{\rm single}$. 
This situation strongly contrasts to that in our ARPES data (Supplementary Figure \ref{WithoutOffsetOP92K}), which clearly show the filling of spectral weight at $E_{\rm F}$ with increasing temperature. The behavior of gap filling can be produced by inputting non-zero $\Gamma_{\rm pair}$ values in Eq.(\ref{self_Norman}). It is demonstrated in 
Supplementary Figure \ref{Simulation}b, where the spectra with several $\Gamma_{\rm pair}$ values are calculated with fixing $\Gamma_{\rm single}$ to zero.
 Importantly, the energy position of spectral peak (${\varepsilon _{{\text{peak}}}}$) dramatically decreases with an increase of the $\Gamma_{\rm pair}$ value as a consequence of the spectral filling around $E_{\rm F}$, and eventually it becomes zero. This simulation indicates that the peak energy underestimates the ``real" energy gap (${\varepsilon _{{\text{peak}}}} < \Delta $) when the magnitude of $\Gamma_{\rm pair}$ becomes large. 
 
Here we should point out that the following self-energy has been used to estimate the magnitude of $\Delta$ by a fitting to the ARPRS spectra in the many previous reports,
\renewcommand {\theequation}{S\arabic{equation}}
\begin{eqnarray} \label{self_simple}
\Sigma ({\bf {k_{\rm F}}},\omega ) \equiv  - i{\Gamma _{{\text{single}}}} + {\Delta ^2}/\omega.
\end{eqnarray}  
This is the same formula as Eq.(\ref{self_Norman}) setting ${\Gamma _{{\text{pair}}}} = 0$, and thus
it always provides one the spectral intensity of zero at $E_{\rm F}$ as discussed above (see Supplementary Figure \ref{Simulation}a). 
We emphasize that the spectral weight around  $E_{\rm F}$ observed in the ARPES data was previously attributed to the spectral broadening coming from the finite energy resolution in ARPES (for example, see the supplementary information of ref.\cite{Shen_BCS}). 
The extremely high energy resolution in our equipment enabled us to reveal 
that the spectral weight at $E_{\rm F}$ is not an artifact 
due to the experimental resolution, but an intrinsic signature of the pair breaking effect.
This signifies that the  Eq.(\ref{self_Norman}) is more suitable than Eq.(\ref{self_simple}) as the model spectral function of cuprates,
and that the effect of $\Gamma_{\rm pair}$ must be taken into account to properly estimate the value of $\Delta$ from the spectral shape.  

In Supplementary Figure \ref{Simulation}c, we simulate a special case setting 
  $\Gamma_{\rm pair} =\Gamma_{\rm single}$ in Eq.(\ref{self_Norman}).
  The self-energy and the corresponding spectral function at $\bf {k_{\rm F}}$ under this condition are derived as follow;
  \renewcommand {\theequation}{S\arabic{equation}}
\begin{eqnarray} \label{BCS_spectrum}
\Sigma ({\bf {k_{\rm F}}},\omega ) =  - i{\Gamma} + {\Delta ^2}/[\omega  + i{\Gamma}],
\nonumber \\ \nonumber \\
A({\bf {k_{\rm F}}},\omega ) = {1 \over 2}\left[ {{\Gamma  \over {{{\left( {\omega  - \Delta } \right)}^2} + {\Gamma ^2}}} + {\Gamma  \over {{{\left( {\omega  + \Delta } \right)}^2} + {\Gamma ^2}}}} \right],
\end{eqnarray}  
where $\Gamma  \equiv {\Gamma _{{\text{single}}}} = {\Gamma _{{\text{pair}}}}$.
This formula, consisting of double Lorentzian functions, is  familiar as the phenomenological BCS spectral function \cite{Schrieffer,JC,Matsui}.
The calculated spectra for various $\Gamma$ values are plotted in Supplementary Figure \ref{Simulation}c. One can confirm a couple of discrepancies in these from our data.  
First, the spectral peak in the ARPRS data has an asymmetric shape about the peak energy, differently from the Lorentzian shape.
Secondly, 
the effect of gap filling in this simulation  is quite sensitive to the magnitude of $\Gamma$, and it is far more drastic than that observed in our data at elevated temperatures.
These lead us to conclude that the Eq.(\ref{BCS_spectrum}) is not a proper formula of spectral function for cuprates either; that is, 
the assumption of $\Gamma_{\rm pair} =\Gamma_{\rm single}$ is not suitable.

We also stress that the Dynes function \cite{Dynes} given by,  
\begin{equation} \label{Dynes} 
{I_{{\rm{DOS}}}}(\omega ) = {\mathop{\rm Re}\nolimits} \left[ {{{\omega  - i\Gamma } \over {\sqrt {{{\left( {\omega  - i\Gamma } \right)}^2} - {\Delta ^2}} }}} \right],
\end{equation} 
is the density-of-states version of Eq.(\ref{BCS_spectrum}) with  $\Gamma  \equiv {\Gamma _{{\text{single}}}} = {\Gamma _{{\text{pair}}}}$, hence it is also not adequate for cuprates.  This is understandable because there is no good physical reason that these two different scattering rates should be equal. 
Our results demonstrate that a proper combination of $\Gamma_{\rm single}$ and $\Gamma_{\rm pair}$ is required to reproduce the data.  

In the previous reports \cite{Dessau_NP,Dessau_PRB,Kondo_MDC}, the spectrum integrated along a selected momentum cut
was utilized to study the gap evolution with temperature and its relation with the electron scattering. 
Here we argue the necessity of investigating the one-particle spectrum, rather than its momentum integration, for such a study. 
In Supplementary Figure \ref{DOSsimulation}, we demonstrate that a momentum-integrated spectrum, ${I_{{\rm{DOS}}}}(\omega ) = \int {A({\bf k},\omega )} d{\bf k}$ for $A({\bf k},\omega )$ of Eq.(\ref{self_Norman}),  is equally sensitive to the $\Gamma_{\rm single}$ value (Supplementary Figs. \ref{DOSsimulation}a and \ref{DOSsimulation}b)
and the $\Gamma_{\rm pair}$ value (Supplementary Figs. \ref{DOSsimulation}c and \ref{DOSsimulation}d); these two scattering rates both fill the spectral intensity around $E_F$, while the gap size $\Delta$ is fixed to a constant value (15 meV).
The spectra of ${I_{{\rm{DOS}}}}(\omega )$ calculated under the condition of ${\Gamma _{{\rm{single}}}} = {\Gamma _{{\rm{pair}}}}$  (Supplementary Figs.   \ref{DOSsimulation}e and  \ref{DOSsimulation}f) reproduce those in Supplementary Figs.  \ref{DOSsimulation}a and \ref{DOSsimulation}b. As naturally expected, only the half values of ${\Gamma _{{\rm{single}}}}$ and  ${\Gamma _{{\rm{pair}}}}$ 
are sufficient to cause the same effect in this case. 
Notably, the obtained curves are  identical to Dynes function in Eq.(\ref{Dynes}) as pointed out above, which is demonstrated for several values of $\Gamma$ in  Supplementary Figure \ref{DOSsimulation}g.  These simulations providing identical results (Supplementary Figs.  \ref{DOSsimulation}a, \ref{DOSsimulation}c, \ref{DOSsimulation}e, and \ref{DOSsimulation}g)  signify that the 
  momentum-integrated spectrum is not capable of extracting the mutual relationship between the two scattering rates (${\Gamma _{{\rm{single}}}}$ and ${\Gamma _{{\rm{pair}}}}$).

In contrast to it,  the effects of  ${\Gamma _{{\rm{single}}}}$ and ${\Gamma _{{\rm{pair}}}}$ on the line shape of the momentum-resolved spectrum $A({\bf k},\omega )$ are independent, 
broadening the width and filling the weight around $E_F$, respectively.
This circumstance is demonstrated in Supplementary Figure \ref{Simulation}, and also
 seen in  Supplementary Figure \ref{DOSsimulation}{\bf h}, at which the spectra at $\bf {k_{\rm F}}$ are extracted from images in Supplementary Figs. \ref{DOSsimulation}{\bf b}, \ref{DOSsimulation}{\bf d}, and \ref{DOSsimulation}{\bf f}. (Spectral colors of red, blue, and green are corresponding to those of thick dashed lines on  the images.) 
Therefore, the  precise investigation for the line shape of one-particle spectra is required 
  to separate the  three important parameters (${\Gamma _{{\rm{single}}}}$,  ${\Gamma _{{\rm{pair}}}}$, and $\Delta$) in   Eq.(\ref{self_Norman}). The main challenge in this task, which has been troubling the ARPES people,  
  is to accomplish both of a sufficient energy resolution and an ultra-high statistics in the ARPES spectra with no noise at a time, which are usually conflicting with each other. We have overcome this difficulty by using a laser ARPES with a unique cold finger (mentioned in "Experimental Method"), which was constructed in our lab. 
We emphasize that the Eq.(\ref{self_Norman}) is a minimal model to reproduce the ARPES data, 
 thus the relation among the three physical parameters, revealed in the present work, provides a key ingredient to formulate the mechanism of the high-$T_{\rm c}$ superconductivity in cuprates. 
\\\\

  {\bf Supplementary Note 5: Signature of a gap existence in the off-nodal, single-peak spectra above $T_{\rm c}$}\\

Here we demonstrate that the off-nodal spectra above $T_{\rm c}$ exhibit a signature of the gap opening even after becoming one-peak structure. 
The bottom panels of Supplementary Figure \ref{OnePeakGap}a and \ref{OnePeakGap}b plot the symmetrized EDCs of OP92K above $T_{\rm c}$ measured at the node ($\phi=0^\circ$) and off the node ($\phi=13.5^\circ$), respectively.
Two peaks seen in the off-nodal spectra merge to a single peak at $T_{\rm one}$=103K (Supplementary Figure \ref{OnePeakGap}a). 
For clarity, the spectra above the $T_{\rm one}$ are indicated with rainbow colors, whereas 
 those at the lower temperatures with gray colors in the both of Supplementary Figure \ref{OnePeakGap}a and \ref{OnePeakGap}a.
 
We find that the single-peak spectra off the node (rainbow curves in Supplementary Figure \ref{OnePeakGap}a) become sharper  
with increasing temperature up to $\sim 135$K ($T_{\rm pair}$) before eventually broadened at higher temperatures.
The anomalous feature is more clearly demonstrated in the upper panel of Supplementary Figure \ref{OnePeakGap}a,
where the spectra below and above the $T_{\rm pair}$ are separated with an offset. 
As seen in the the magnified insets, the spectral tail shifts to lower energies in the range of $T<T_{\rm pair}$ (a  magenta arrow), 
while it to the higher energies in $T>T_{\rm pair}$ (a black arrow).
It strongly contrasts to the case at the node (Supplementary Figure \ref{OnePeakGap}b); the spectral tail keeps shifting toward higher energies at elevated temperatures in the both temperature ranges (black arrows). 
We summarize these results in Supplementary Figure  \ref{OnePeakGap}c, by plotting the spectral width $\Delta\varepsilon$ (HWHM) at and off the node as a function of temperature. The anomalous decrease of $\Delta\varepsilon$ is seen only for the data off the node (orange circles). 
As discussed in the main paper, a plausible explanation for it 
is that the off-nodal spectra have small energy gaps, even though they exhibit single peak shapes, and they become sharper up to the  temperate ($T_{\rm pair}$) at which the energy gap completely closes. 
\\\\

  {\bf Supplementary Note 6: Spectral fitting under the assumption of $\Delta (T) \equiv {\varepsilon _{{\text{peak}}}}(T)$}\\

In the main paper, we claim that the energy of spectral peak underestimates the energy gap (${\varepsilon _{{\text{peak}}}} < \Delta $) close to the node at high temperatures, at which the spectral peak width becomes larger than the magnitude of the energy gap. 
Here we perform a spectral fitting to our ARPES data with Eq.(\ref{self_Norman}) [same as Eq.(1) in the main paper], assuming that  
these two values are identical [$\Delta (T) \equiv {\varepsilon _{{\text{peak}}}}(T)$,  upper panel of Supplementary Figure \ref{UnrealisticOP92K}a (OP92K) and Supplementary Figure \ref{UnrealisticOD72K}a (OD72K)], 
and demonstrate that  the extracted  parameter has an unrealistic behavior, reflecting the irrelevance of $\Delta (T) \equiv {\varepsilon _{{\text{peak}}}}(T)$. 
The middle and bottom panels of  Supplementary Figure \ref{UnrealisticOP92K}a  plot  
the  $\Gamma_{\rm single}(T)$ and $\Gamma_{\rm pair}(T)$ of OP92K, respectively, obtained by the fitting to the spectra in the momentum region, where the Fermi arc was previously claimed to appear at $T_{\rm c}$ (orange arrow in the insets of Supplementary Figure \ref{UnrealisticOP92K}b). 
 For a comparison, we also plot in Supplementary Figure \ref{UnrealisticOP92K}b (OP92K) the corresponding results for  $\Delta (T)$ of the BCS-like gap function, which are presented in the main paper (Fig. 5a). 
  The difference between the two fitting-results is very clear; 
the former (Supplementary Figure \ref{UnrealisticOP92K}a) shows an abnormal upturn in the  ${\Gamma _{{\text{single}}}}(T)$ curves on cooling in the gapped momentum region, and it contrasts to 
the latter result  (Supplementary Figure \ref{UnrealisticOP92K}b)  showing almost identical curves in ${\Gamma _{{\text{single}}}}(T)$ for all
$\phi$s. 
In  Supplementary Figure \ref{UnrealisticOD72K}, we present the same data set as in Supplementary Figure \ref{UnrealisticOP92K}, but for OD72K. 
Similarly to the case of OP92K, an abnormal upturn is obtained in the  ${\Gamma _{{\text{single}}}}(T)$ when $\Delta (T) \equiv {\varepsilon _{{\text{peak}}}}(T)$ (the middle panel of Supplementary Figure \ref{UnrealisticOD72K}a), and it is corrected by setting $\Delta (T)$ to be the BCS-like gap function  (the middle panel of Supplementary Figure \ref{UnrealisticOD72K}b).  

The  behavior of the abnormal upturn is interpreted as follows. 
 An energy gap is still open above $T_{\rm c}$ even though the spectrum has a single peak, and thus the  spectral peak width overestimates the scattering rate.
 Since the degree of the overestimation gets higher with an increase of the energy gap on cooling, the upturn in the ${\Gamma _{{\text{single}}}}(T)$ appears. 
The upturn  seems to be more enhanced with getting away from the node (see middle panel of Supplementary Figure \ref{UnrealisticOP92K}a and Supplementary Figure \ref{UnrealisticOD72K}a).
 This is also expected as the overestimation becomes more serious toward the antinode with a larger energy gap. 
This scenario is further validated by the fact that the upturns start around $T \sim 130$K and $T \sim 90$K for OP92K and OD72K, respectively, which are consistent with the onset temperatures of pair formation determined in the main paper. 

One might think that the cuprates are known to have 
a significantly anisotropic scattering mechanism, 
thus the situation in Supplementary Figure \ref{UnrealisticOP92K}a (OP92K) and Supplementary Figure \ref{UnrealisticOD72K}a  (OD72K) with the strong momentum variation could be intrinsic. 
To argue against this, here we demonstrate that even the optimally doped Bi2212, in fact, has an isotropic scattering mechanism around the node. 
Supplementary Figures \ref{IsotropicScattering}a and \ref{IsotropicScattering}b show the spectra at various $\bf {k_{\rm F}}$ points  over a wide $\phi$ angle  measured at the lowest temperature ($T$=10K) and above the pairing temperature ($T$=150K), respectively. 
At these temperatures, the value of ${\Gamma _{{\text{single}}}}$ can be  estimated simply from the peak width of spectra,
since the magnitude of  ${\Gamma _{{\text{pair}}}}$ is negligible at 10K, and 
the second term in Eq.(\ref{self_Norman}) with ${\Gamma _{{\text{pair}}}}$  is irrelevant
above the pairing temperature.
The values of ${\Gamma _{{\text{single}}}}$  estimated from the peak width of spectra in Supplementary Figure \ref{IsotropicScattering}a and \ref{IsotropicScattering}b
are plotted in  Supplementary Figure \ref{IsotropicScattering}d with blue and red circles, respectively.
The magnitude is almost constant over a wide $\phi$ centered at the node for both the temperatures, 
which indicates that the scattering mechanism near the node is isotropic. 
This strongly supports  our assertion that the anisotropic behavior of 
${\Gamma _{{\text{single}}}}$ with an upturn on cooling is an artifact. 
Beyond  $\left| \phi  \right| \approx 15^\circ$, the scattering rate abruptly increases. 
This signifies the evolution of the competing pseudogap, which becomes dominant around the antinode \cite{Kondo_competition,Kondo_pair,Fujita,Kohsaka}.  
\\\\

  {\bf Supplementary Note 7: Fitting curves reproducing the ARPES spectra}\\

In the main paper,  
we demonstrate the spectral fitting with a minimal model of Eq.(\ref{self_Norman}) to the data of OP92K at  $\phi=13.5^\circ$ (Fig.5c), as a typical example.
Here we present all of the fitting results for the ARPES spectra of OP92K and OD72K, 
which provide the physical parameters plotted in  Fig. 5a (Supplementary Figure \ref{FittingOP92K}a) and Fig. 5b (Supplementary Figure \ref{FittingOD72K}a), respectively. 
Supplementary Figure \ref{FittingOP92K}b and \ref{FittingOD72K}b shows the ARPES spectra (black curves) and fitting results (red curves) for OP92K and OD72K, respectively. 
The data are almost perfectly reproduced by the fitting curves.  
  \\\\
  
  {\bf Supplementary Note 8: The pairing signature detected in the nodal $\Gamma_{\rm single}(T)$}\\
  
The single-particle scattering rate ($\Gamma_{\rm single}$) at the node is extracted simply from the spectral width,
thus it can be precisely determined without any models.
Supplementary Figure \ref{NodalGamma}a and \ref{NodalGamma}b  plots the temperature dependence of EDC width $\Delta\varepsilon$ at the nodal $\bf {k_{\rm F}}$ point of OP92K and OD72K, respectively (same as $\Gamma_{\rm single} (T)$ at $\phi=0^\circ$ in Figs. 5a and 5b).
We find that the curves of $\Delta\varepsilon (T)$
deviate from the $T$-linear behavior ($\Delta {\varepsilon _{{\rm{linear}}}}(T)$)
in the both samples. It is demonstrated more clearly in the bottom panels by plotting the amount of divination [$\Delta {\varepsilon _{{\rm{linear}}}}(T) - \Delta {\varepsilon _{{\rm{data}}}}(T)$]. 
The onset temperatures of the deviation  are estimated to be $\sim135$K and $\sim89$K for OP92K and OD72K, respectively (yellow circles).
The suppression of the scattering rate should be tied to the electron pairing. We found that the onset temperatures are 
indeed coincident with the $T_{\rm pair}$ values (gap opening temperatures) estimated from the off-nodal spectra in the main paper.
  \\\\
  
  {\bf Supplementary Note 9: Fitting results with a different scheme}\\
  
The model spectral function of  Eq.(\ref{self_Norman}) [same as Eq.(1) in the main paper] has three parameters of the energy gap ($\Delta$), the single-particle scattering rate ($\Gamma_{\rm single}$), 
and the pair breaking rate ($\Gamma_{\rm pair}$).
In the main paper, we assume the $\Delta$ to be BCS-type gap function with the onset temperature much higher than $T_{\rm c}$,
and extract the other two parameters of $\Gamma_{\rm single}$ and $\Gamma_{\rm pair}$ by fitting the model function to the data.
Here we perform a new analysis with a different fitting scheme to justify the BCS-type gap function assumed.
In the new analysis, we use the nodal $\Gamma_{\rm single}(T)$, which can be determined without models,  for the fitting of the off-nodal spectra,
and set the other two parameters of $\Delta$ and $\Gamma_{\rm pair}$ free.
The underlying idea for it is that the scattering mechanism is rather isotropic near the node.
To make the fitting realistic, we slightly shifted the nodal $\Gamma_{\rm single}(T)$ in the magnitude.
The fitting results  (OP92K, $\phi=11.7^\circ$) are demonstrated in Supplementary Figure  \ref{AnotherFitting}a.
The fitting spectra (red curves) reproduce the data (black curves) reasonably well.
The extracted $\Delta$ and $\Gamma_{\rm pair}$ also more or less agree to the results presented in the main paper,
while these values above $\sim 120$K are not properly extracted owing to the small magnitudes of $\Delta$. 
This result justifies that the $\Delta(T)$ in cuprates has 
the BCS-type gap function with an onset higher than $T_{\rm c}$, and validates the requirement of pair breaking rate ($\Gamma_{\rm pair}$) to reproduce the ARPES spectra.
\\\\\\\\\

%\newpage

%%%%%%%%%%%%%%%%%%%%%%
\begin{figure*}  
\includegraphics[width=4.5in]{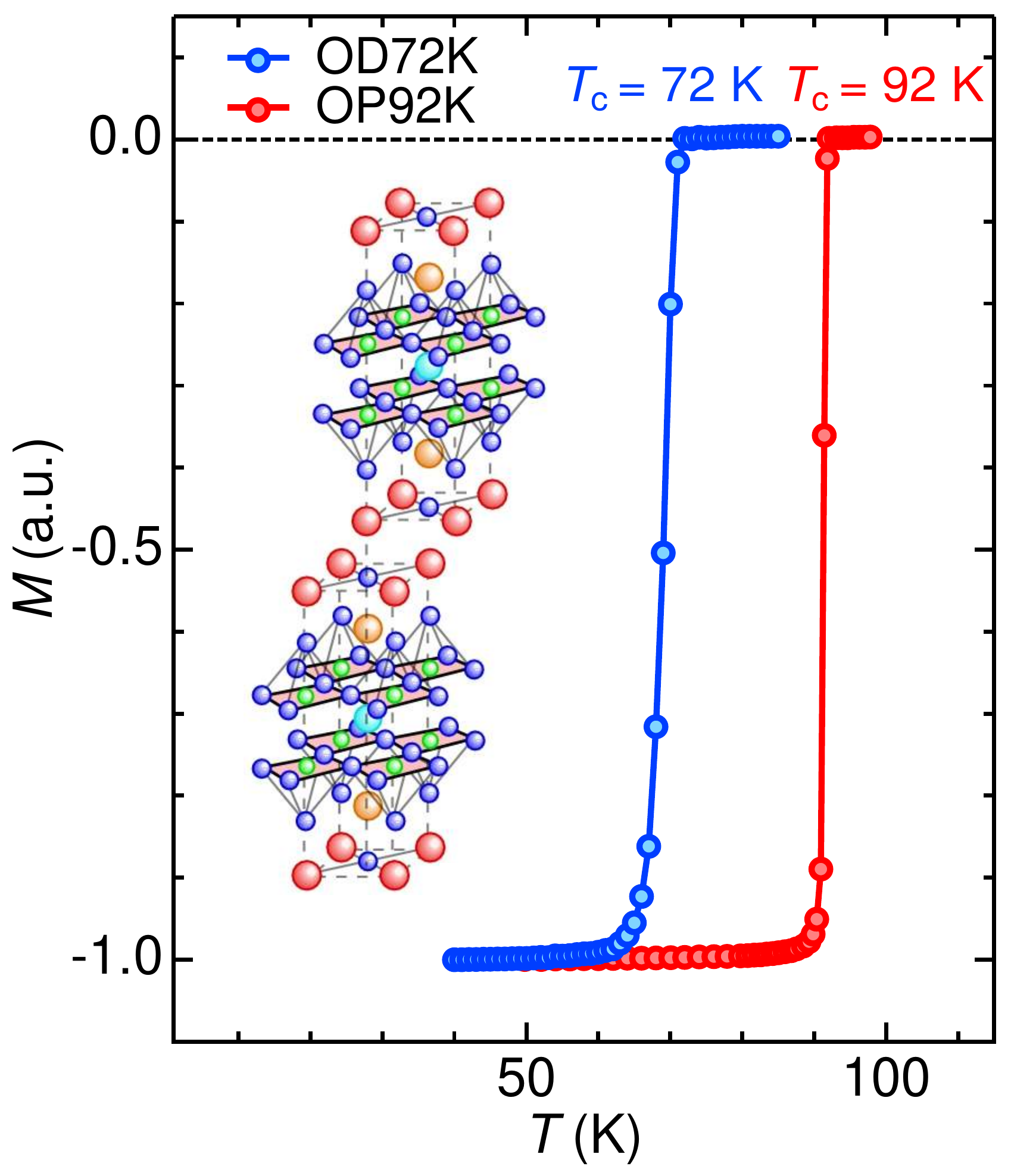}
\caption{Magnetic susceptibility for  single crystals of optimally doped Bi$_2$Sr$_2$CaCu$_2$O$_{8+\delta}$ (OP92K) and overdoped (Bi,Pb)$_2$Sr$_2$CaCu$_2$O$_{8+\delta}$ (OD72K) with an onset $T_{\rm c}$ of 92K and 72K, respectively, used for the ARPES measurements.  
The crystal structure of Bi2212 is drawn in the inset. }
\label{SQUID}
\end{figure*}

%\clearpage
% \newpage

%%%%%%%%%%%%%%%%%%%%%%
\begin{figure*}  
\includegraphics[width=6.0in]{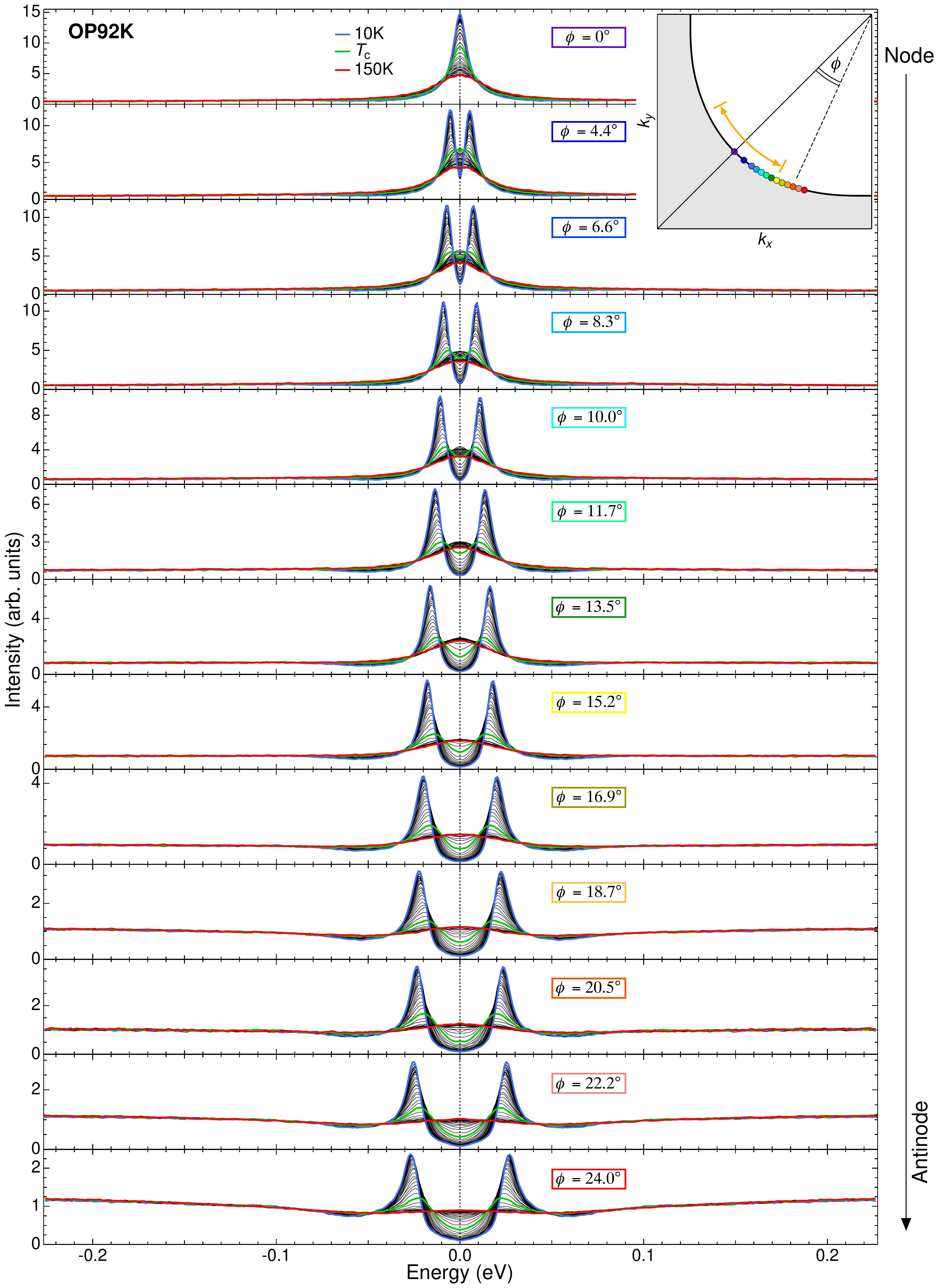}
\caption{Temperature evolution of ARPES spectra (symmetrized EDCs) at various $\bf {k_{\rm F}}$s for OP92K (color circles in the inset). 
The  corresponding $\phi$ angles (defined in the inset)  are described in each panel. 
The orange arrow in the inset indicates the momentum region where the gapless Fermi arc was previously claimed to emerge at $T_{\rm c}$.}
\label{WithoutOffsetOP92K}
\end{figure*}

%\clearpage
% \newpage

 %%%%%%%%%%%%%%%%%%%%%%
\begin{figure*} 
\includegraphics[width=6.3in]{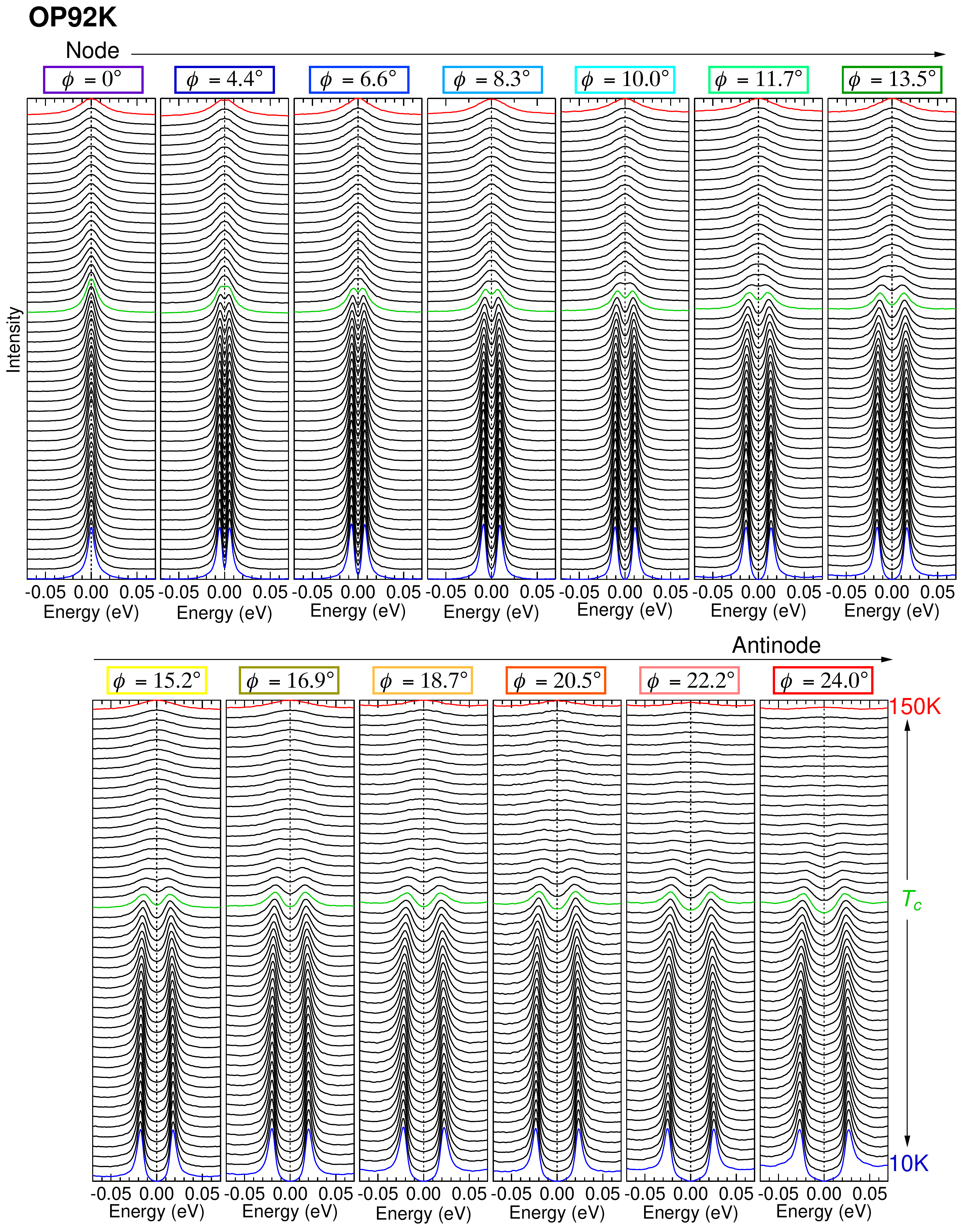}
\caption{Temperature evolution of ARPES spectra (symmetrized EDCs) at various $\bf {k_{\rm F}}$s for OP92K: the same data as in Supplementary Figure \ref{WithoutOffsetOP92K}, but plotted with an offset. 
The top left panel plots the data at the node ($\phi  = 0^\circ$). Toward the bottom right, the measured $\bf {k_{\rm F}}$ approaches the antinode. 
The blue, green, and red curves in each panel are the spectra at 10K, $T_{\rm c}$ (=92K), and 150K, respectively.}
\label{WithOffsetOP92K}
\end{figure*}

%\clearpage
% \newpage

 %%%%%%%%%%%%%%%%%%%%%%
\begin{figure*} \label{}
\includegraphics[width=5.2 in]{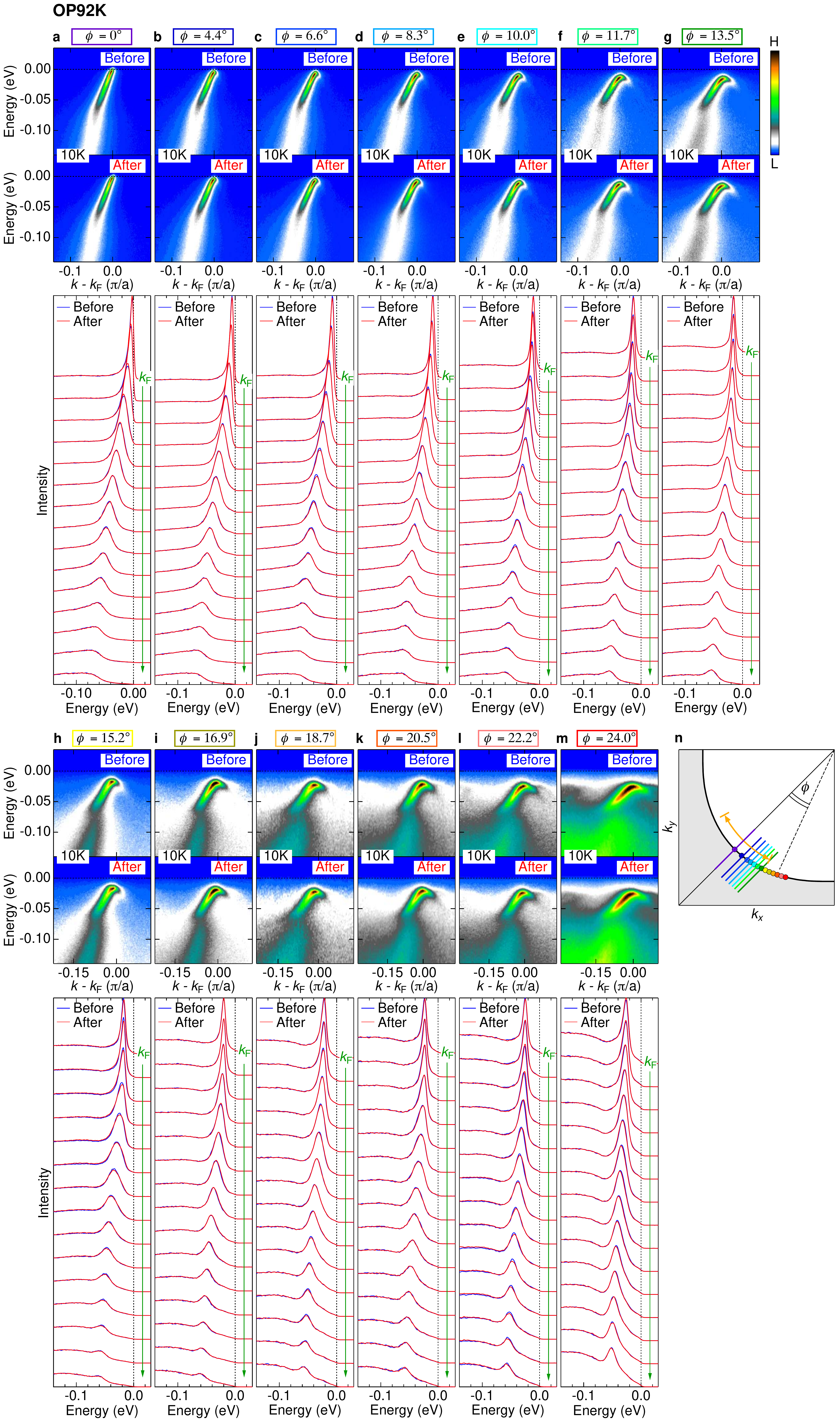}
\caption{
Aging check of the sample surface after temperature scan. The upper left panels in {\bf a} show the data along the node ($\phi  = 0^\circ$). Toward the bottom right in {\bf m}, the measured momentum approaches the antinode.  {\bf n}, Fermi surface with the measured momentum cuts and $\bf {k_{\rm F}}$ points. 
} 
\label{AgingCheck}
\end{figure*}

%\clearpage
% \newpage

 %%%%%%%%%%%%%%%%%%%%%%
\begin{figure*} \label{}
\includegraphics[width=6.1 in]{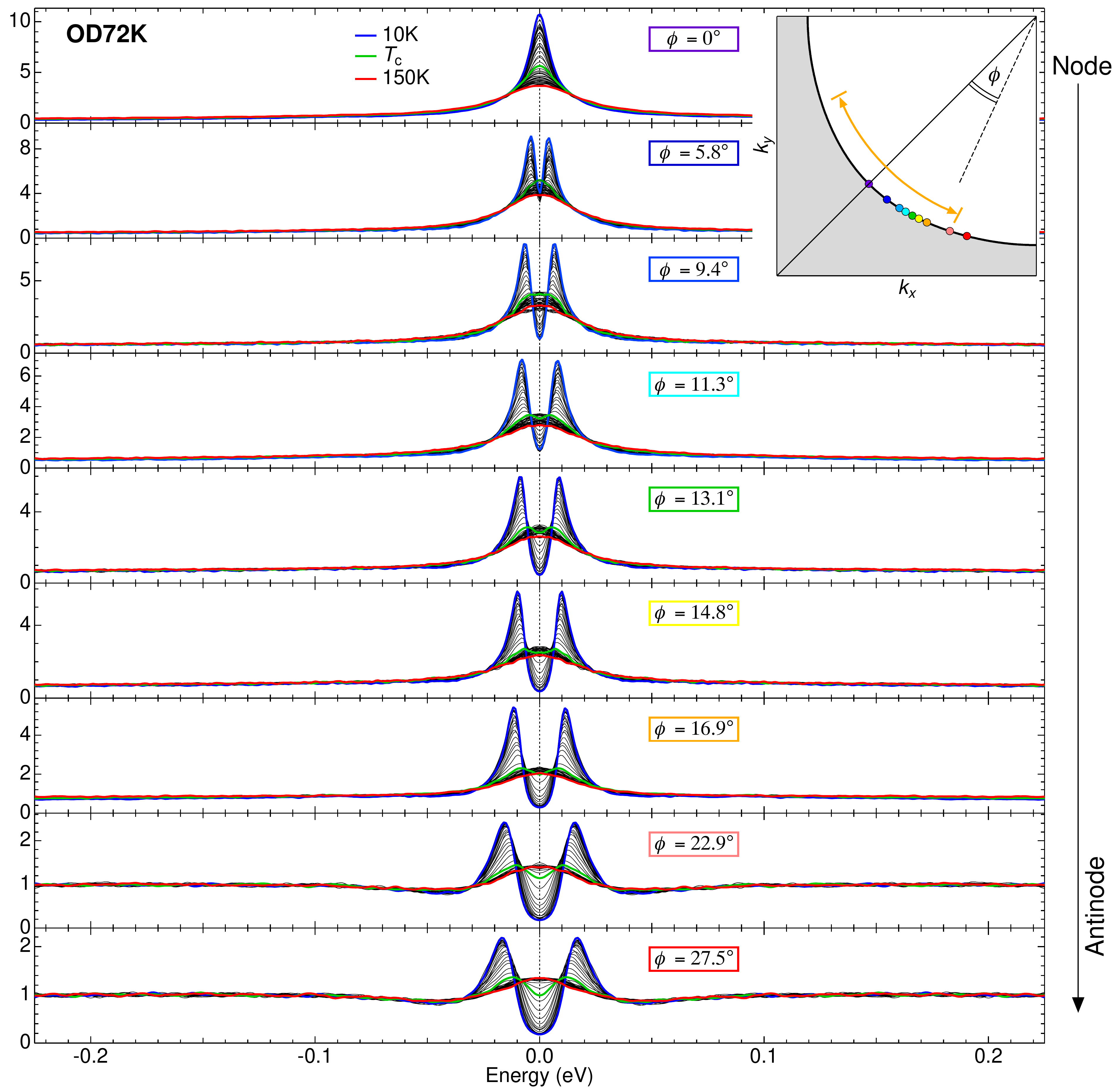}
\caption{Temperature evolution of ARPES spectra (symmetrized EDCs) at various $\bf {k_{\rm F}}$s for OD72K (color circles in the inset). 
The  corresponding $\phi$ angles (defined in the inset)  are described in each panel. 
The orange arrow in the inset indicates the momentum region where the gapless Fermi arc was previously claimed to emerge at $T_{\rm c}$.}
\label{WithoutOffsetOD72K}
\end{figure*}

%\clearpage
% \newpage

 %%%%%%%%%%%%%%%%%%%%%%
\begin{figure*} 
\includegraphics[width=6.3in]{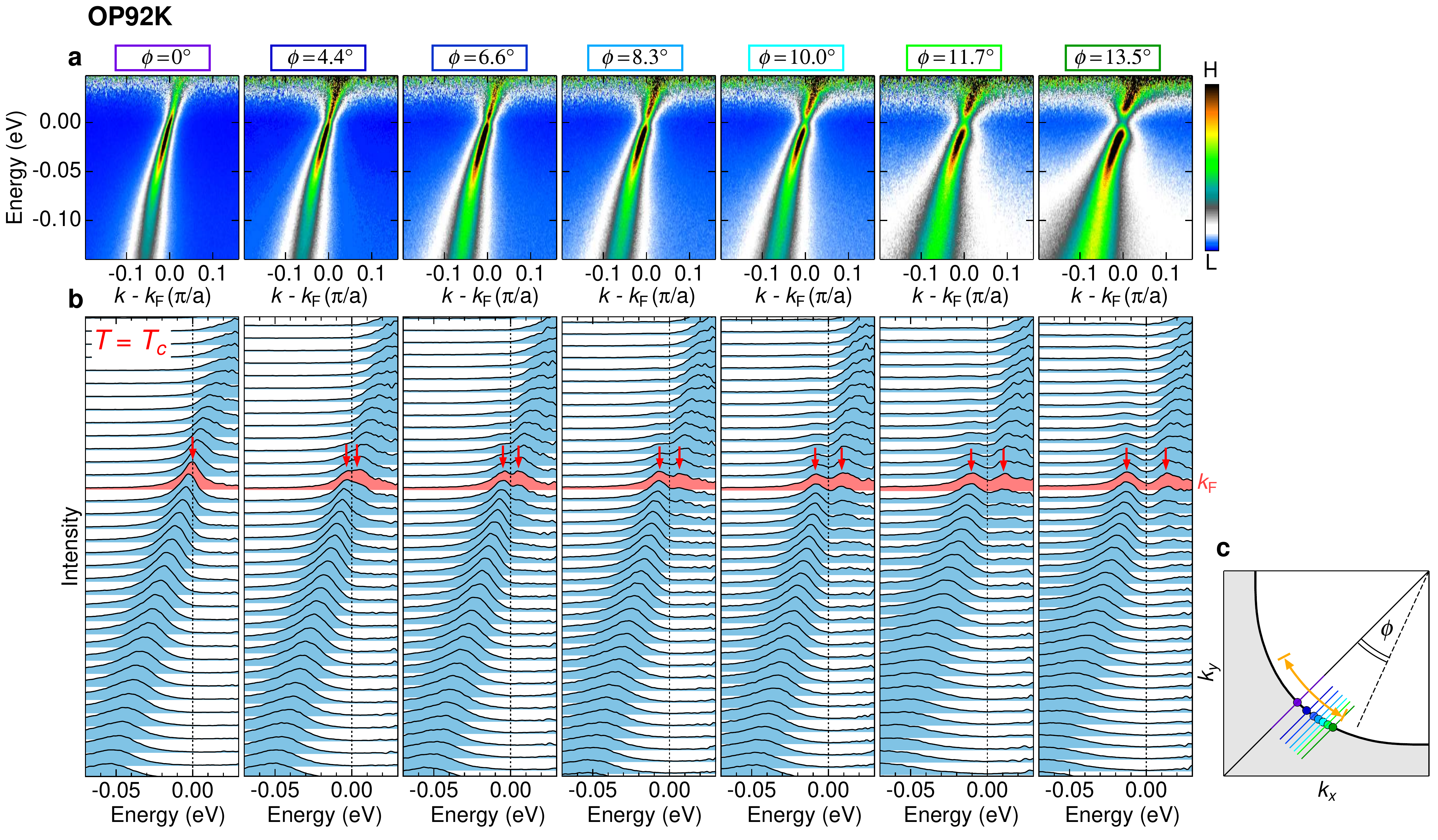}
\caption{Evidence for the absence of the Fermi arc at $T_{\rm c}$.  {\bf a},  Fermi-function divided band dispersions at $T_{\rm c}$ along several momentum cuts (color lines in {\bf c}).  Same data as Fig.3a in the main paper.
{\bf b},  The corresponding EDCs.  The spectra at $\bf {k_{\rm F}}$s  (color circles in {\bf c}) are painted with red. Peak positions are indicated with red arrows. 
{\bf c}, The Fermi surface. 
The orange arrow indicates the momentum region where the gapless Fermi arc was previously claimed to emerge at $T_{\rm c}$.
}
\label{Pointnode}
\end{figure*}

%\clearpage
% \newpage

%%%%%%%%%%%%%%%%%%%%%%
\begin{figure*} 
\includegraphics[width=5.3in]{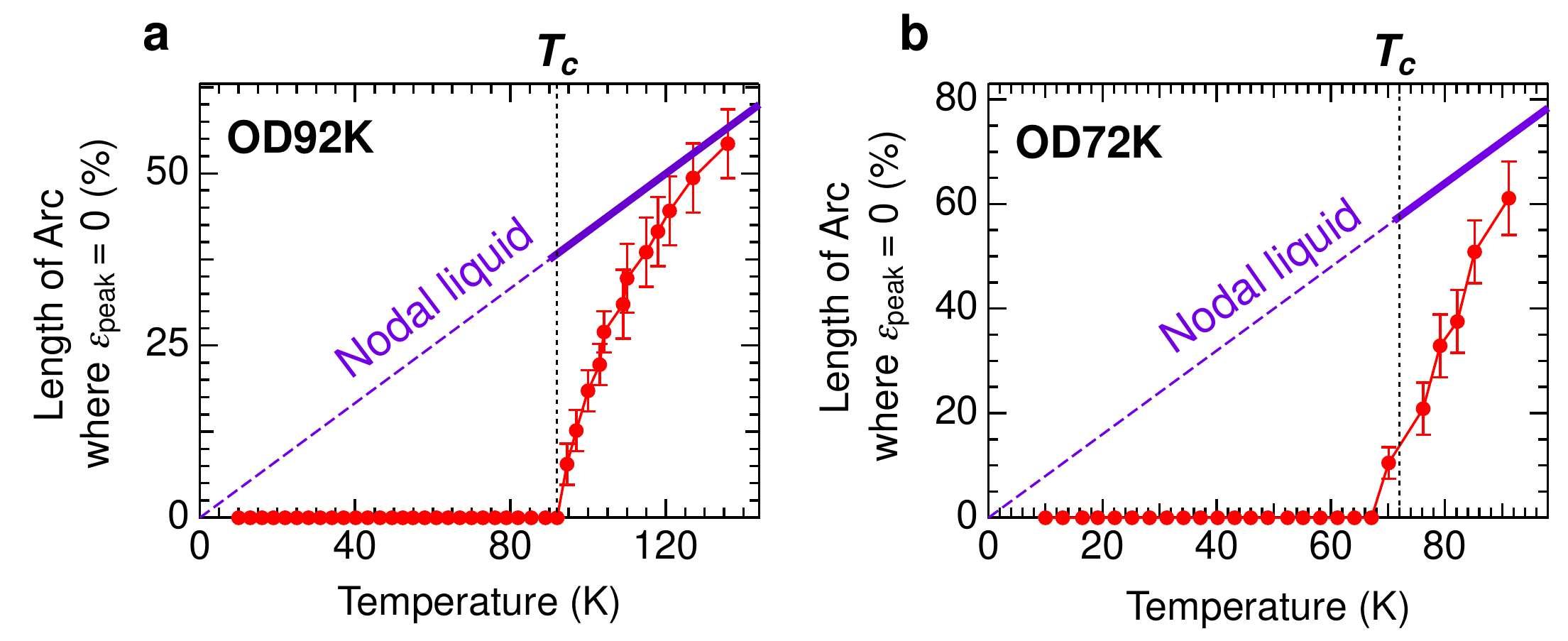}
\caption{The ``artificial" Fermi arc. The length of arc where the symmetrized EDCs have a single-peak (${\varepsilon _{{\rm{peak}}}}$=0) for OP92K ({\bf a}) and OD72K ({\bf b}).
The nodal liquid behavior, which has a $T$-linear fashion and is extrapolated to zero at $T=0$, is superimposed in each panel.
}
\label{ArcLength}
\end{figure*}

%\clearpage
% \newpage

 %%%%%%%%%%%%%%%%%%%%%%
\begin{figure*} \label{}
\includegraphics[width=6.3in]{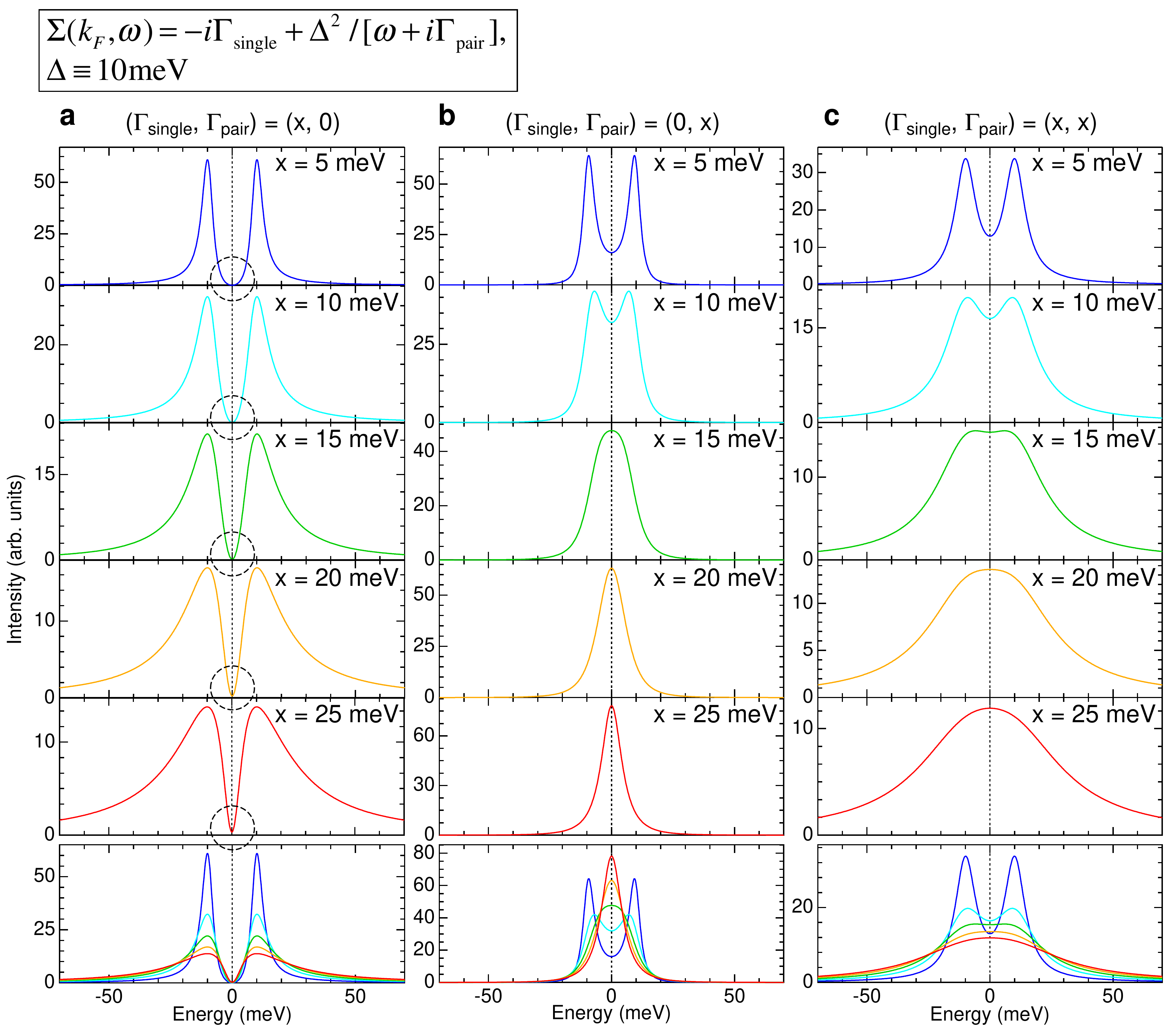}
\caption{Simulation of spectral function in Eq.(\ref{self_Norman}) for examining the effect of  
 two different scattering rates (${\Gamma _{{\text{single}}}}$ and ${\Gamma _{{\text{pair}}}}$) on the spectral shape.  
In all the curves presented, the gap magnitude is set to 10meV ($\Delta  \equiv 10$ meV). 
 {\bf a}, The ${\Gamma _{{\text{single}}}}$ dependence when ${\Gamma _{{\text{pair}}}}  \equiv 0$meV.
 {\bf b}, The ${\Gamma _{{\text{pair}}}}$ dependence when ${\Gamma _{{\text{single}}}} \equiv 0$meV.  
 {\bf c}, The spectral shape for various values of ${\Gamma _{{\text{single}}}}$ and ${\Gamma _{{\text{pair}}}}$, which  
are set to be equal. In each bottom panel of {\bf a}, {\bf b}, and {\bf c}, all spectra in the upper panels are superimposed. 
While all the curves plotted are convoluted with a gaussian that has the width of the experimental energy resolution ($\Delta \varepsilon  = 1.4$meV), the $\Delta \varepsilon$ value is so small that the difference in shape from original curves is negligible.} 
\label{Simulation}
\end{figure*}

%\clearpage
% \newpage

%%%%%%%%%%%%%%%%%%%%%%
\begin{figure*} \label{}
\includegraphics[width=6.3in]{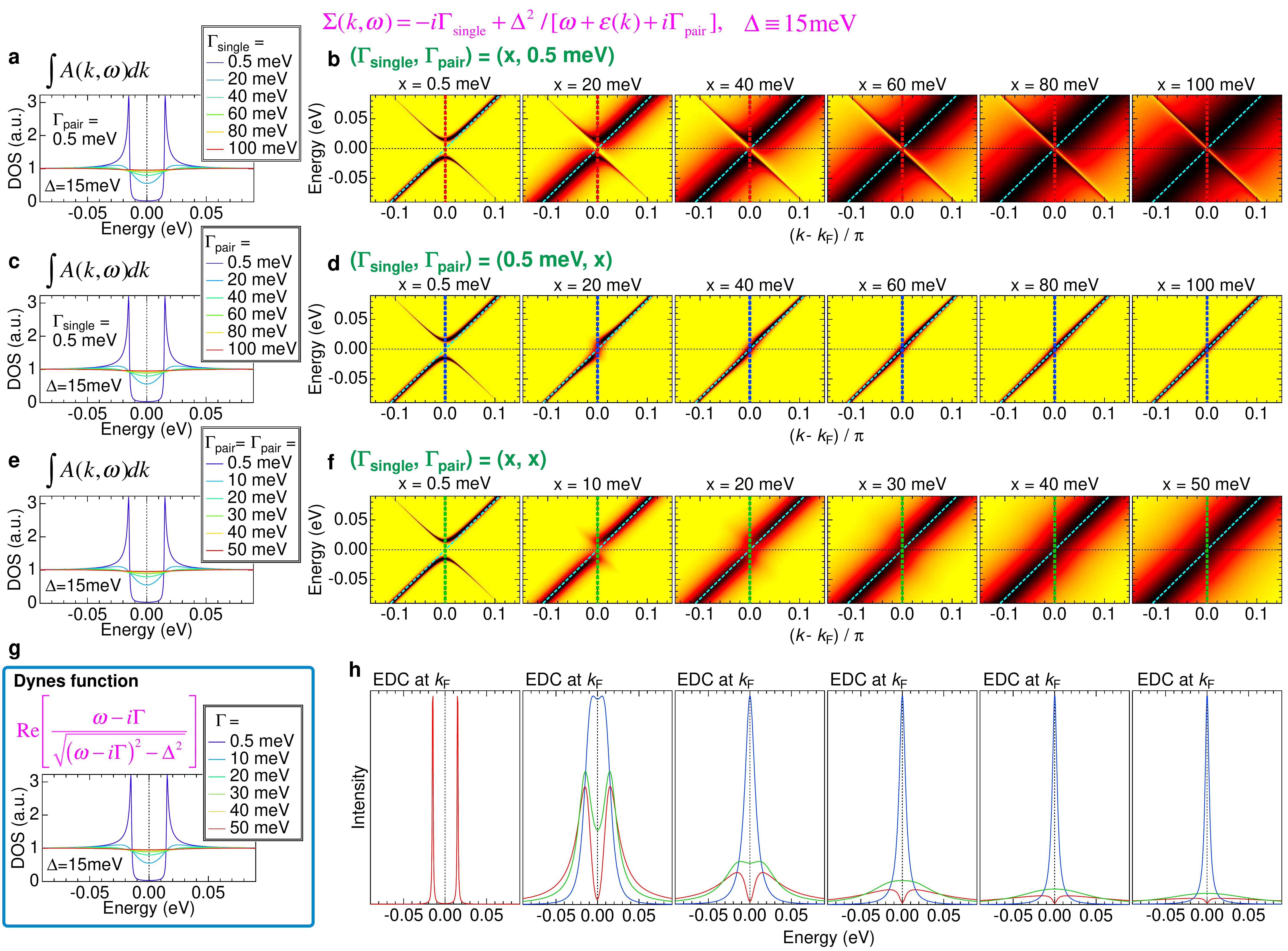}
\caption{Simulations for examining the effect of ${\Gamma _{{\text{single}}}}$ and ${\Gamma _{{\text{pair}}}}$ on the momentum-integrated spectra. The magnitude of energy gap $\Delta$ is fixed to 15 meV for all the cases.
{\bf a, c, e},  Momentum integrated spectra, ${I_{{\rm{DOS}}}}(\omega ) = \int {A({\bf k},\omega )} dk$, 
 for the dispersion images in {\bf b, d, f}, respectively.
{\bf b, d, f}, Dispersion images for one-particle spectral function, $A({\bf k},\omega )$, in Eq.(\ref{self_Norman}) with a linear $\varepsilon (k) $ (light blue dashed lines).
$\Gamma_{\rm single}$ and $\Gamma_{\rm pair}$ dependences are studied in ({\bf a, b}) and ({\bf c, d}), respectively. The case under the condition of ${\Gamma _{{\text{single}}}} = {\Gamma _{{\text{pair}}}}$ is simulated in ({\bf e, f}). {\bf g}, The spectra of Dynes function with several values of  $\Gamma$. 
{\bf h}, Spectra at $\bf {k_{\rm F}}$, $A(\bf {k_{\rm F}},\omega )$, extracted from images of {\bf b, d, f}.
 Each color (red, blue, and green) in curves corresponds to that of thick dashed line added in the images of  {\bf b, d, f}.
}
\label{DOSsimulation}
\end{figure*}

%\clearpage
% \newpage
 
 %%%%%%%%%%%%%%%%%%%%%%
\begin{figure*} \label{}
\includegraphics[width=6.3in]{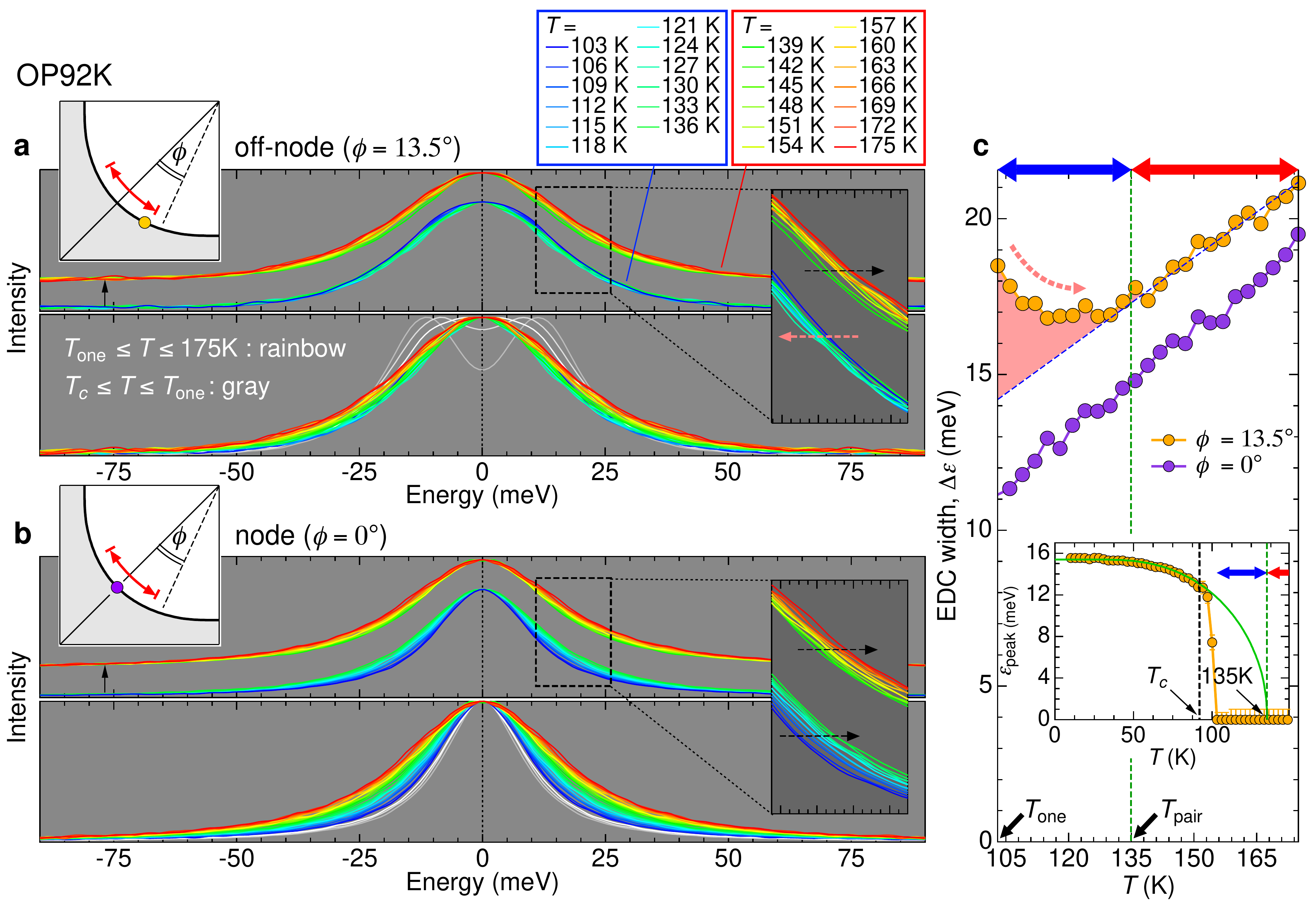}
\caption{Signature for the gap existence in the single-peak spectra off the node. 
Symmetrized EDCs of OP92K at $T>T_{\rm c}$ measured off the node (bottom panel of {\bf a}) and at the node (bottom panel of {\bf b}).
The $\bf {k_{\rm F}}$ points are marked with circles in the inset of  {\bf a} ($\phi=13.5^\circ$) and {\bf b} ($\phi=0^\circ$). 
Two peaks in the off-nodal spectra merge to one peak at 103K ($T_{\rm one}$).
{\bf c}, The spectral width ($\Delta \varepsilon$) at $T>T_{\rm one}$ are plotted as a function of temperature. 
The off-nodal $\Delta \varepsilon (T)$ decreases with temperature and reaches the minimum at $\sim 135$K ($T_{\rm pair}$) before it eventually increases at elevated temperatures. 
The anomalous behavior contrasts to the nodal $\Delta \varepsilon (T)$, which monotonically increases over the whole range of temperatures.  
In the upper panels of {\bf a} and {\bf b}, the spectra above $T_{\rm one}$ with a single-peak (rainbow colors) are extracted. 
In addition, the spectra at $T>T_{\rm pair}$ and $T<T_{\rm pair}$ are separated with an offset (black arrow). 
The insets show the magnified areas marked with the dashed squares. 
The black and magenta dashed arrows indicate the energy shifts of spectral tails with increasing temperature. 
The inset of {\bf c} plots the energy positions of symmetrized EDCs ($\varepsilon_{\rm peak}$) as a function of temperature. 
The green curve indicates the BCS-type gap function with the onset of 135K, which fits well the $\varepsilon_{\rm peak}(T)$ at the low temperatures.
}
\label{OnePeakGap}
\label{fig1}
\end{figure*}

%\clearpage
% \newpage
 
 %%%%%%%%%%%%%%%%%%%%%%
\begin{figure*} \label{}
\includegraphics[width=6.3in]{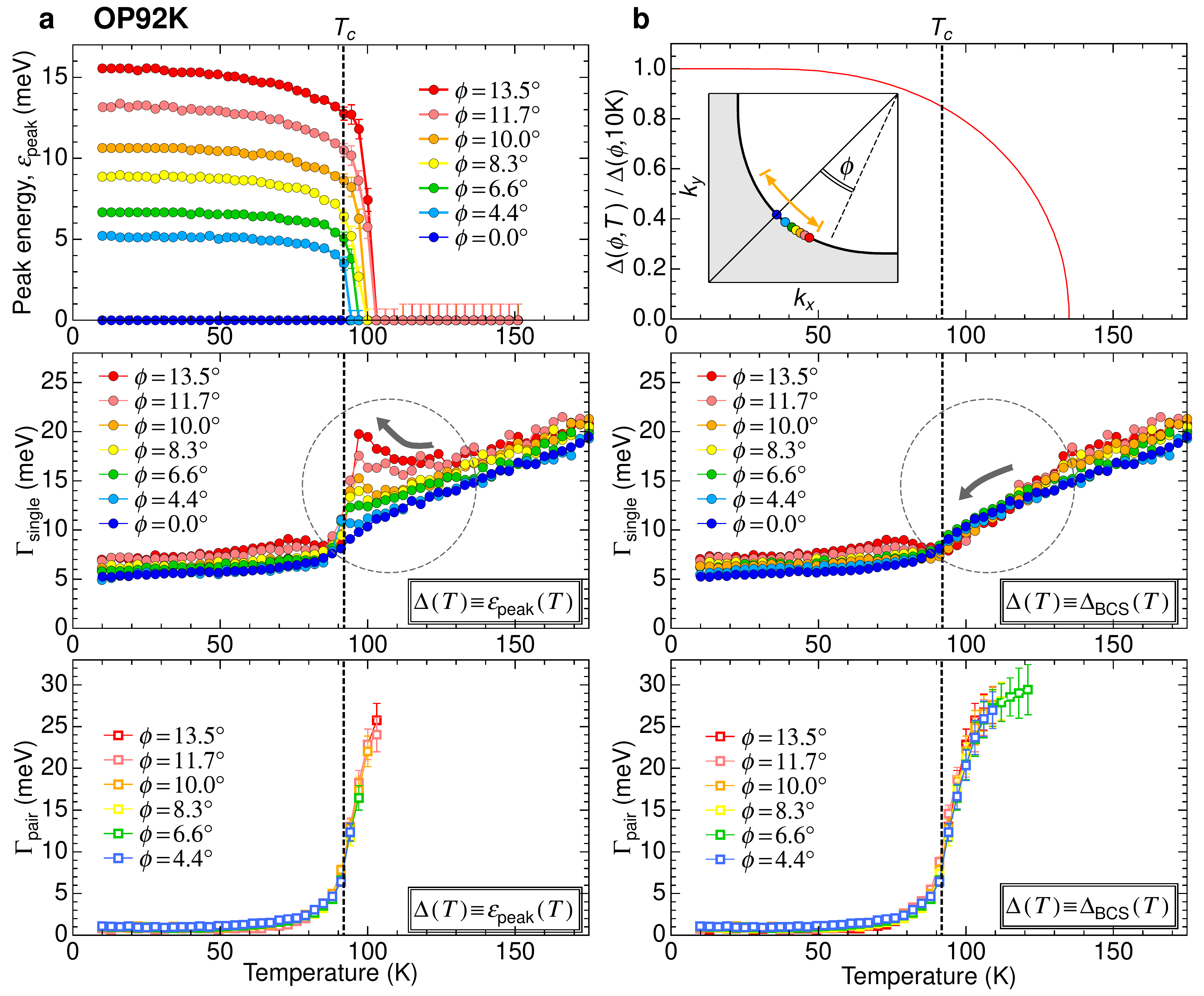}
\caption{Comparison of fitting results with Eq.(\ref{self_Norman}) for OP92K data
between two different settings to the energy gap,  $\Delta $.  {\bf a}, The $\Delta $ is defined to be equal to the peak position of spectra [$\Delta (T) \equiv {\varepsilon _{{\text{peak}}}}(T)$], shown in the top panel. 
{\bf b}, The $\Delta$ is defined to have a BCS-type gap function with an onset at $T_{\rm pair}$=135K, shown in the top panel. 
The  middle and bottom panels plot 
the obtained fitting parameters,  ${\Gamma _{{\text{single}}}}$  and ${\Gamma _{{\text{pair}}}}$, at several $\bf {k_{\rm F}}$ points in the momentum region, where the gapless Fermi arc was previously claimed to emerge at $T_{\rm c}$ (see circles and an arrow in the inset of {\bf b}). 
The values of $\Gamma_{\rm pair}$ at high temperatures are not plotted, since 
the spectral shape is not sensitive to the $\Gamma_{\rm pair}$ when $\Delta$ is small or zero, and thus it is impossible to precisely determine the value.
}
\label{UnrealisticOP92K}
\label{fig1}
\end{figure*}

%\clearpage
% \newpage
 
%%%%%%%%%%%%%%%%%%%%%%
\begin{figure*} \label{}
\includegraphics[width=6.3in]{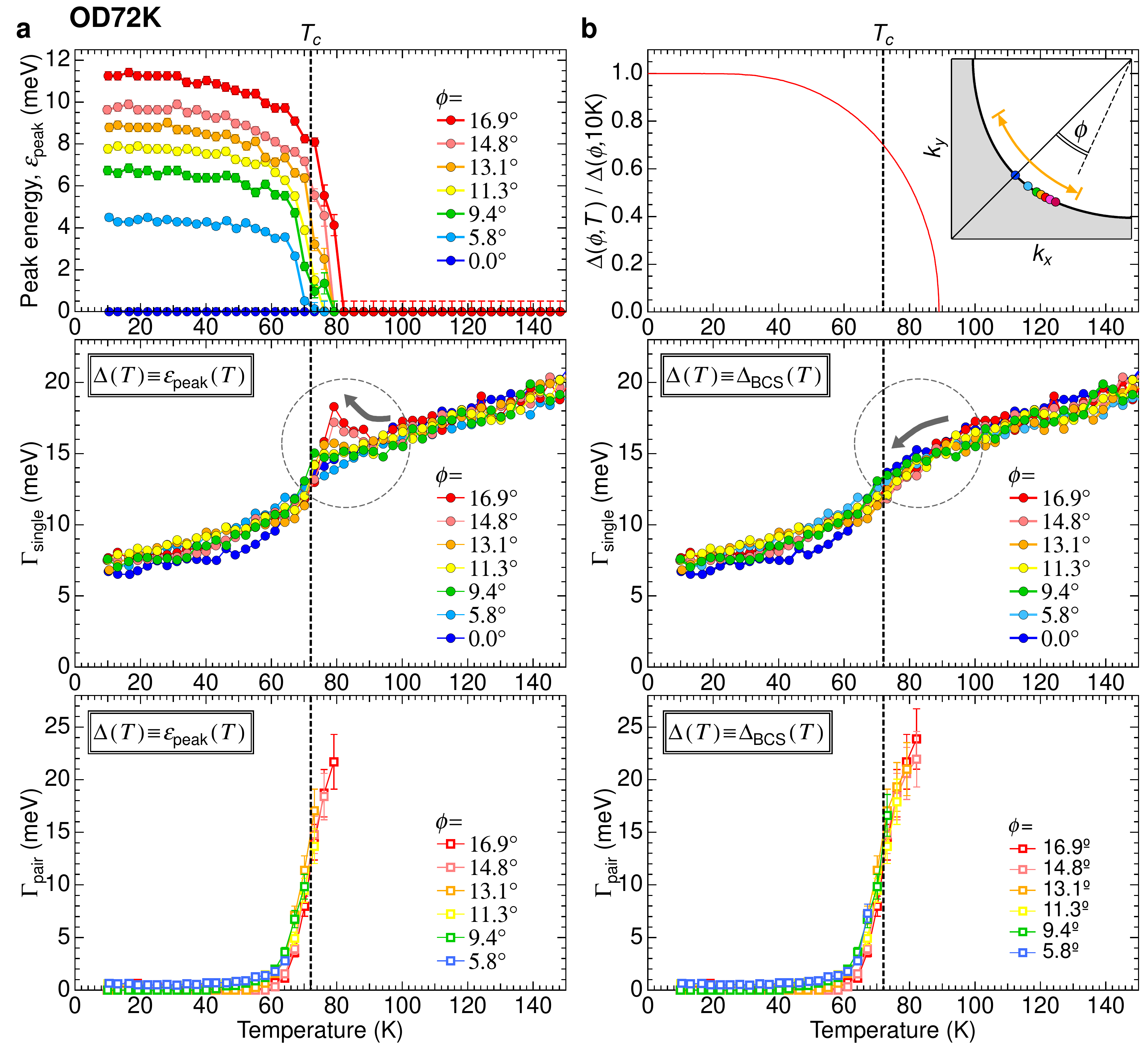}
\caption{Comparison of fitting results with Eq.(\ref{self_Norman}) for OD72K data
between two different settings to the energy gap,  $\Delta $.  {\bf a}, The $\Delta $ is defined to be equal to the peak position of spectra [$\Delta (T) \equiv {\varepsilon _{{\text{peak}}}}(T)$], shown in the top panel. 
{\bf b}, The $\Delta$ is defined to have a BCS-type gap function with an onset at $T_{\rm pair}$=89K, shown in the top panel. 
The  middle and bottom panels plot 
the obtained fitting parameters,  ${\Gamma _{{\text{single}}}}$  and ${\Gamma _{{\text{pair}}}}$, at several $\bf {k_{\rm F}}$ points in the momentum region, where the gapless Fermi arc was previously claimed to emerge at $T_{\rm c}$ (see circles and an arrow in the inset of {\bf b}). 
The values of $\Gamma_{\rm pair}$ at high temperatures are not plotted, since 
the spectral shape is not sensitive to the $\Gamma_{\rm pair}$ when $\Delta$ is small or zero, and thus it is impossible to precisely determine the value.}
\label{UnrealisticOD72K}
\label{fig1}
\end{figure*}

%\clearpage
% \newpage
 
 %%%%%%%%%%%%%%%%%%%%%%
\begin{figure*} \label{}
\includegraphics[width=5.0in]{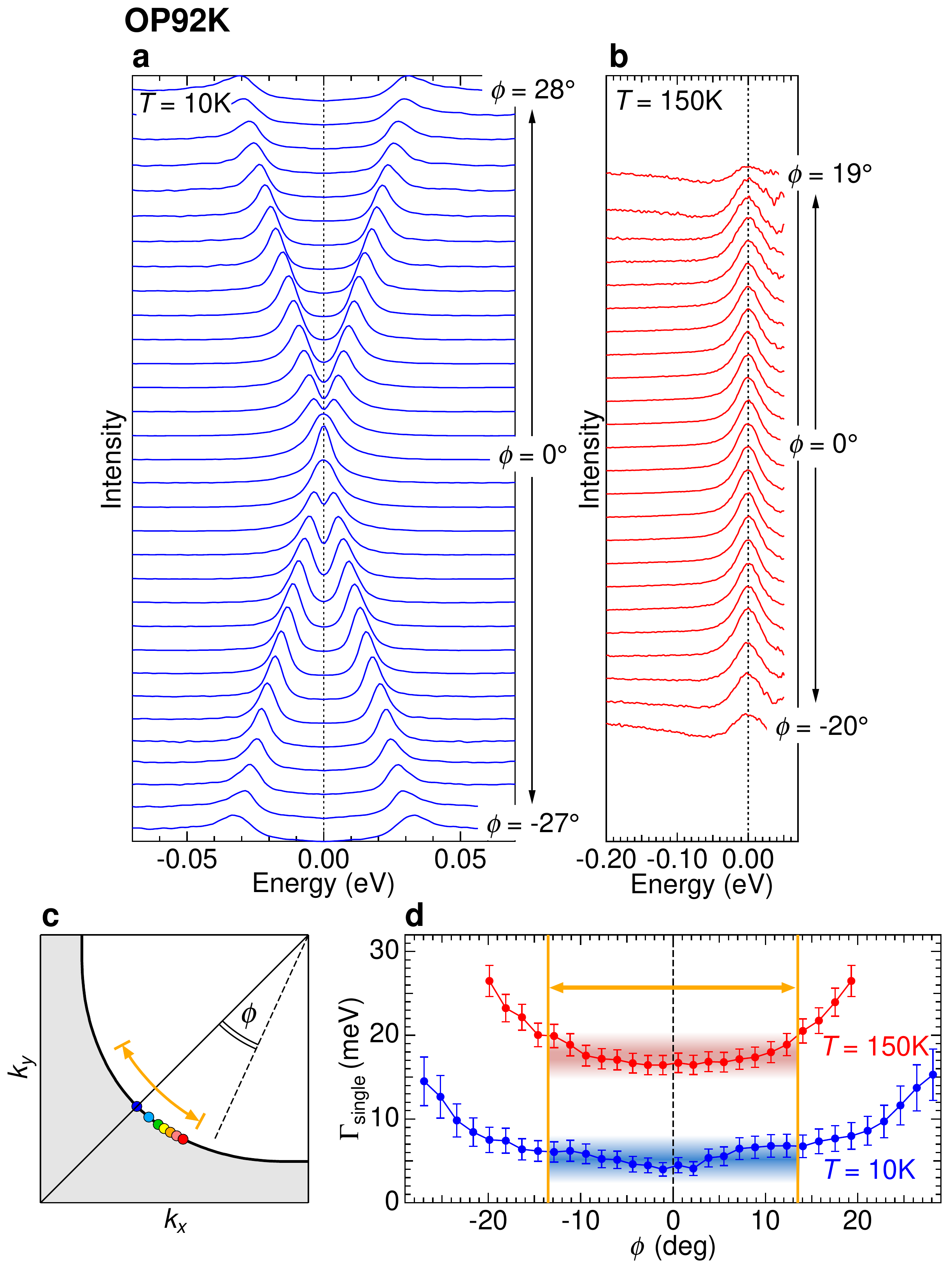}
\caption{Isotropic scattering mechanism around the node. 
{\bf a}, {\bf b},  ARPES spectra of OP92K measured at various $\bf {k_{\rm F}}$ points over a wide $\phi$ angle measured far below $T_{\rm c}$ ($T$=10K) ({\bf a}) and   above $T_{\rm pair}$ ($T$=150K) ({\bf b}). In order to remove the effect of Fermi cut-off, the curves in {\bf a} are symmetrized about the Fermi level, and the ones in {\bf b} are divided by the Fermi function at 150K.  {\bf c}, The Fermi surface.  The $\bf {k_{\rm F}}$ points studied in Fig.5a and Supplementary Figure \ref{FittingOP92K} are marked by colored circles. 
{\bf d}, The peak width of spectra (or ${\Gamma _{{\text{single}}}}$) in {\bf a} and {\bf b}. 
The orange arrows in {\bf c} and {\bf d} indicate the momentum region where the gapless Fermi arc was previously claimed to emerge at $T_{\rm c}$. The almost isotropic ${\Gamma _{{\text{single}}}}$ is obtained in this momentum region. 
}
\label{IsotropicScattering}
\label{fig1}
\end{figure*}

%\clearpage
% \newpage
 
 %%%%%%%%%%%%%%%%%%%%%%
\begin{figure*} \label{}
\includegraphics[width=5.5in]{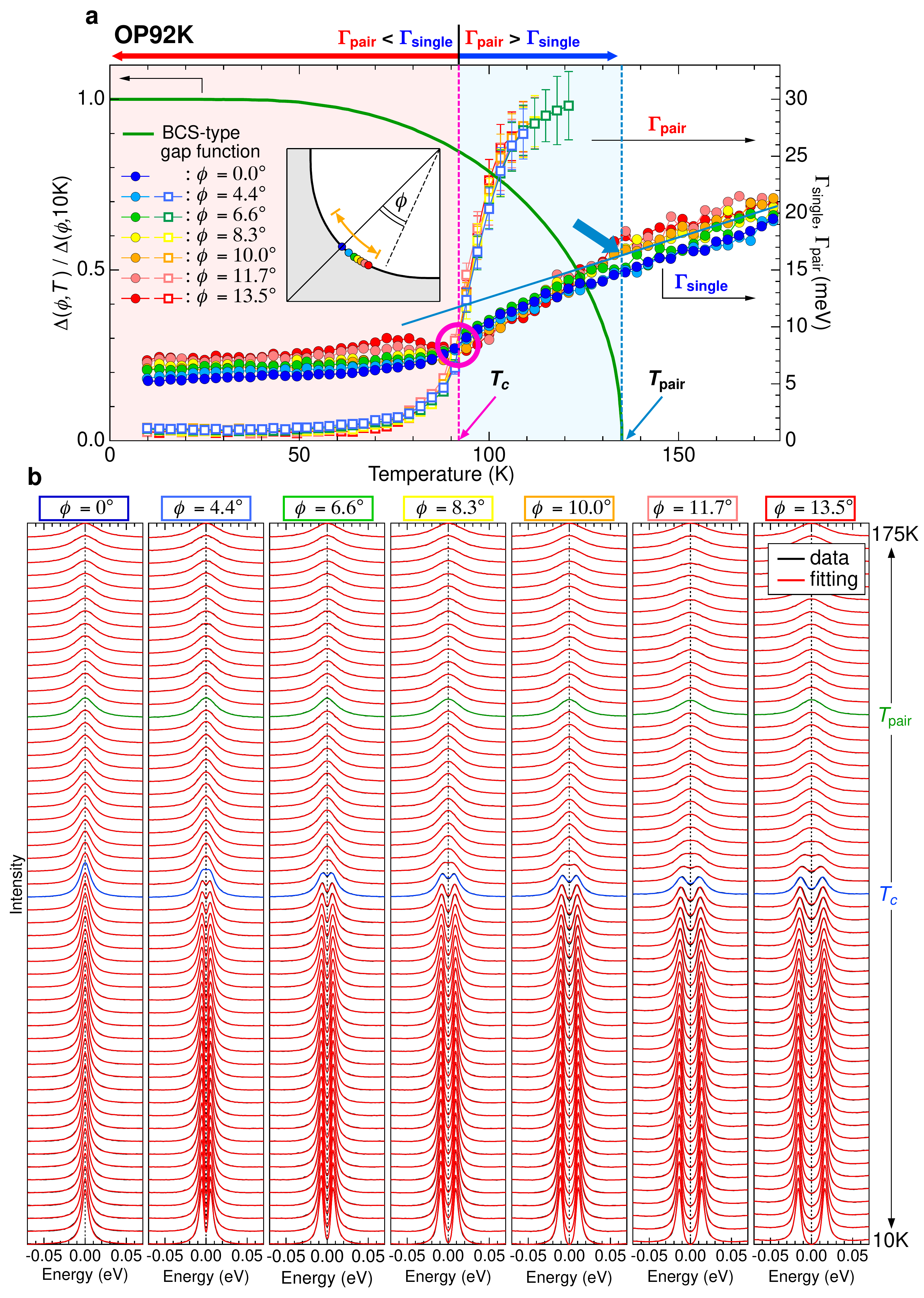}
\caption{Fitting results for OP92K data with the model function of Eq.(\ref{self_Norman}). {\bf a}, Same panel as Fig. 5a in the main paper: single-particle scattering rate ($\Gamma_{\rm single}$) and  pair breaking rate ($\Gamma_{\rm pair}$) extracted by setting the energy gap ($\Delta$) to be 
the BCS-type gap function (a green curve) with the onset of 135K. 
 {\bf b}, ARPES spectra (black curves) and fitting results (red curves) are overlapped.  Fermi angle ($\phi$) is described in the top of each panel, and the corresponding $\bf {k_{\rm F}}$ point is marked with a circle in the inset of {\bf a}. 
 We added a small background linear in energy ($\propto \left| \omega  \right|$) to the fitting function $A({k_F},\omega )$ for $\phi=11.7^{\circ}$ and $13.5^{\circ}$, 
in order to properly extract the scattering rates.
}
\label{FittingOP92K}
\label{fig1}
\end{figure*}

%\clearpage
% \newpage
 
%%%%%%%%%%%%%%%%%%%%%%
\begin{figure*} \label{}
\includegraphics[width=5.5in]{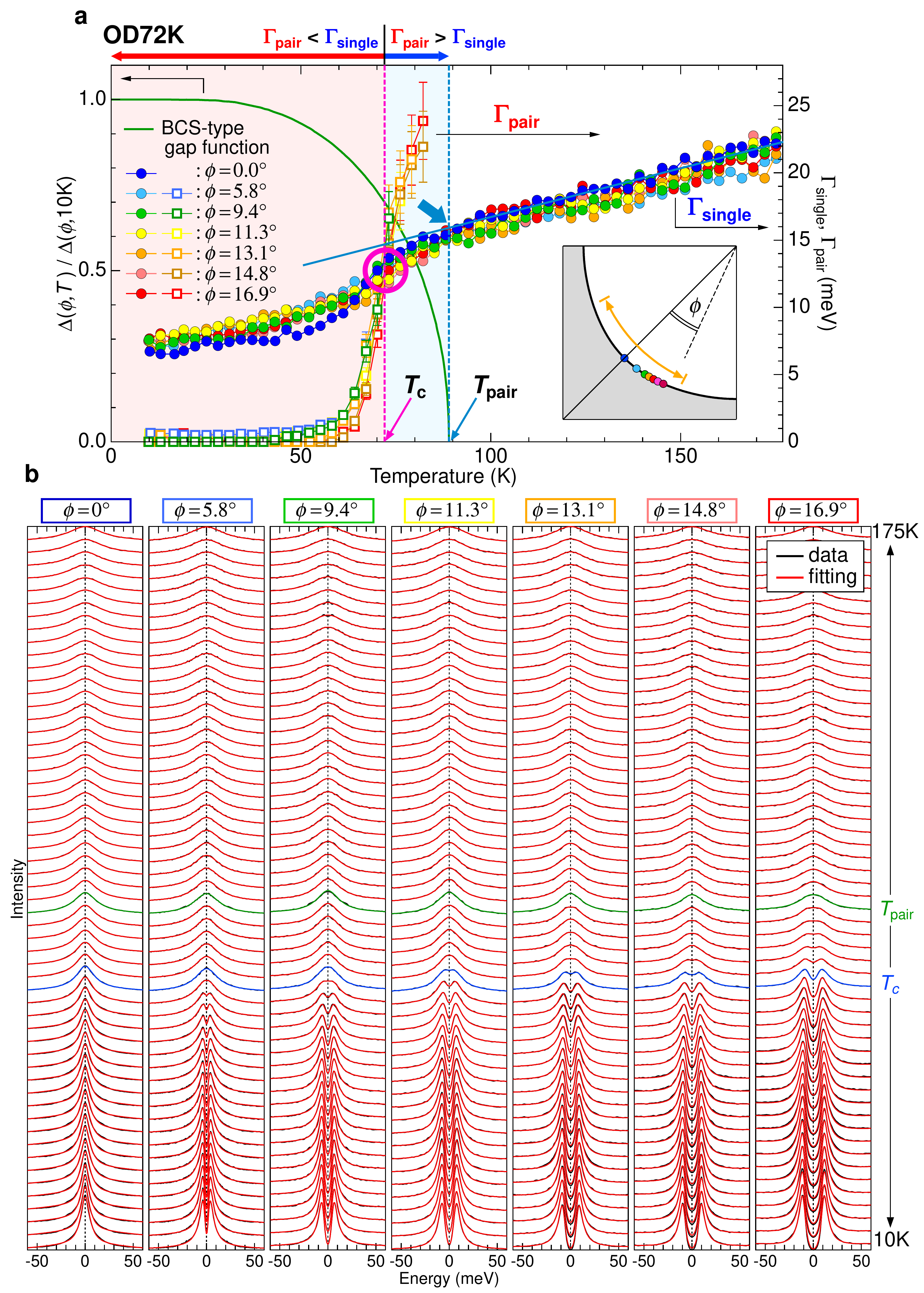}
\caption{Fitting results for OD72K data with the model function of Eq.(\ref{self_Norman}). {\bf a}, Same panel as Fig. 5b in the main paper: single-particle scattering rate ($\Gamma_{\rm single}$) and  pair breaking rate ($\Gamma_{\rm pair}$) extracted by setting the energy gap ($\Delta$) to be 
the BCS-type gap function (a green curve) with the onset of 89K. 
 {\bf b}, ARPES spectra (black curves) and fitting results (red curves) are overlapped.  Fermi angle ($\phi$) is described in the top of each panel, and the corresponding $\bf {k_{\rm F}}$ point is marked with a circle in the inset of {\bf a}. 
  We added a small background linear in energy ($\propto \left| \omega  \right|$) to the fitting function $A({k_F},\omega )$ for $\phi=14.8^{\circ}$ and $16.9^{\circ}$, 
in order to properly extract the scattering rates.
}
\label{FittingOD72K}
\label{fig1}
\end{figure*}

%\clearpage
% \newpage
 
 %%%%%%%%%%%%%%%%%%%%%%
\begin{figure*} \label{}
\includegraphics[width=6.3in]{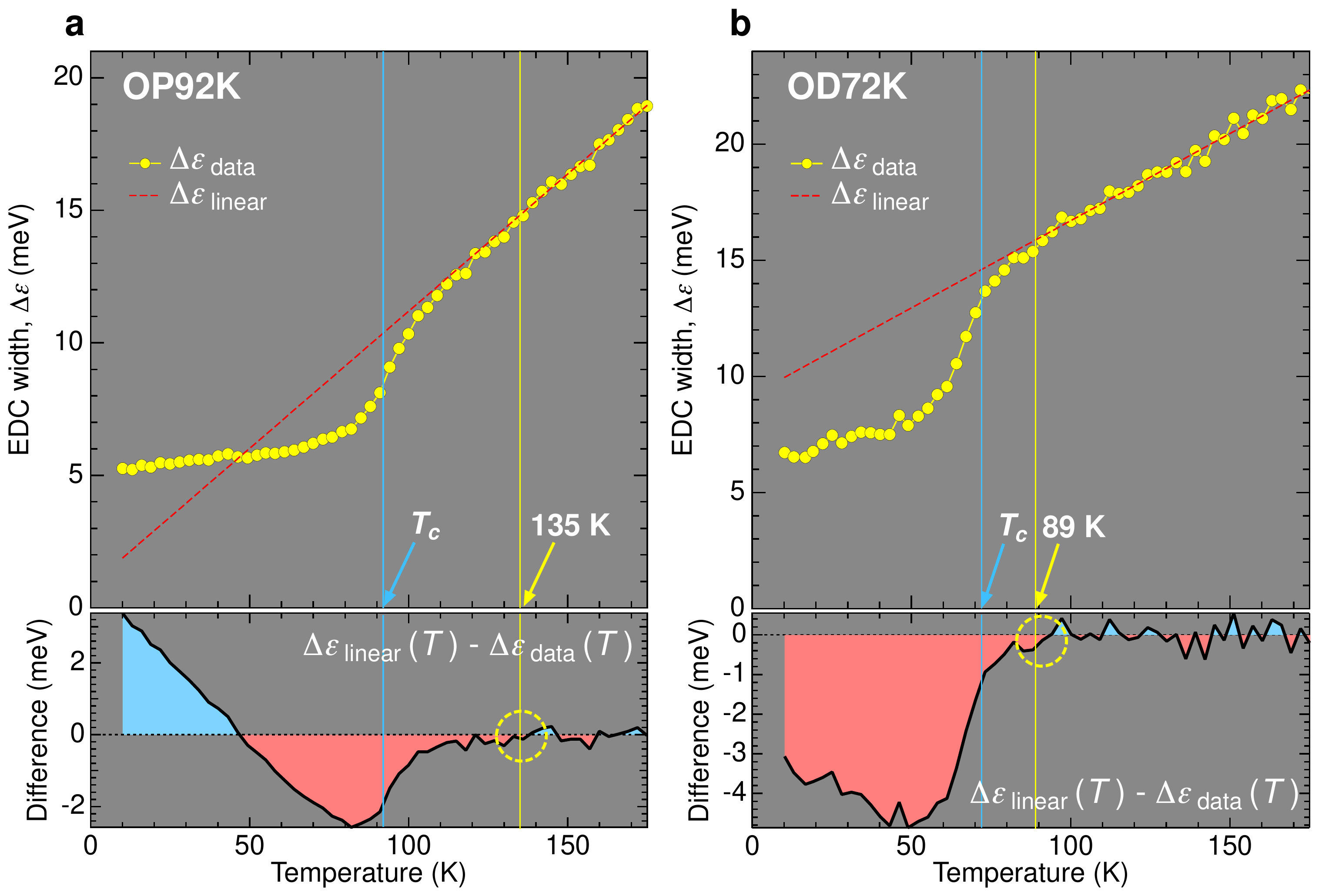}
\caption{
Signature of the pairing detected in the nodal spectra. 
Temperature evolution of peak width ($\Delta \varepsilon_{\rm data}$) of the symmetrized EDCs for OP92K ({\bf a}) and OD72K ({\bf b}) are plotted.
In the bottom of each panel, the deviation of the $\Delta \varepsilon_{\rm data}(T)$ from the $T$-linear behaviors (red dashed lines) are estimated.
The onset temperatures of the deviation are marked by yellow circles. 
The $T_{\rm c}$ and $T_{\rm pair}$ determined in the main paper are indicated by a light-blue and yellow arrow, respectively.}
\label{NodalGamma}
\label{fig1}
\end{figure*}

%\clearpage
% \newpage

 %%%%%%%%%%%%%%%%%%%%%%
\begin{figure*} \label{}
\includegraphics[width=4.3in]{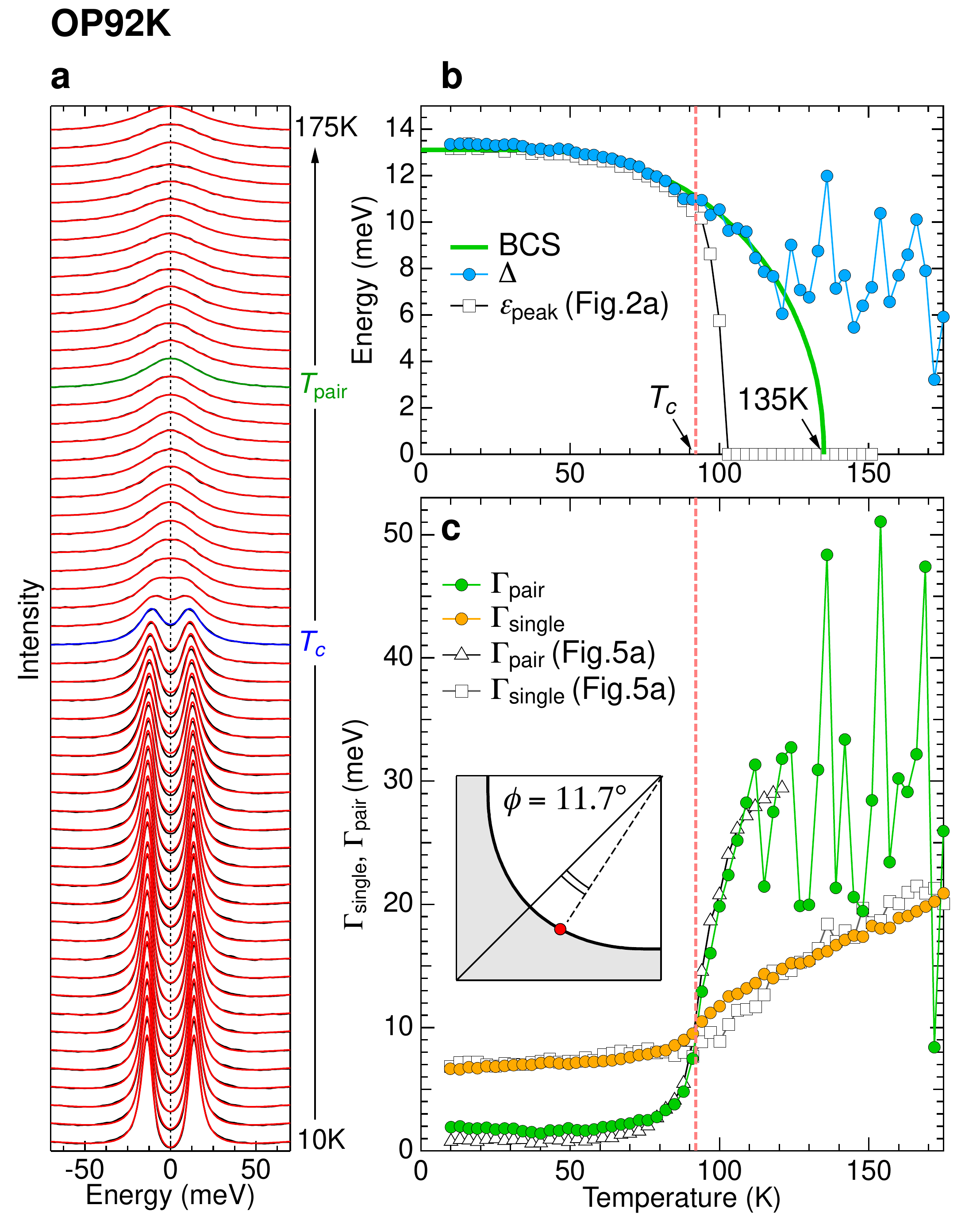}
\caption{Fitting results with a different scheme from that used in the main paper; here we input model values to $\Gamma _{\rm single}(T)$ in  Eq.(\ref{self_Norman}) and extract the other two parameters of $\Delta(T)$ and $\Gamma _{\rm pair}(T)$. 
 {\bf a}, The off-nodal ARPES spectra of OP92K (black curves) and the fitting results (red curves) at $\phi=11.7^\circ$ (red circle in the inset of {\bf c}). 
We used the nodal $\Gamma _{\rm single}(T)$ slightly shifted in the magnitude  (orange circles in {\bf c}), 
and added a small background linear in energy ($\propto \left| \omega  \right|$) to the $A({k_F},\omega )$ in  Eq.(\ref{self_Norman}) 
for a realistic fitting to extract the other two parameters ($\Delta$ and $\Gamma _{\rm pair}$). 
{\bf b}, The obtained $\Delta (T)$ are plotted (light blue circles). For comparison, the BCS-type gap function with the onset of 135K (green curve) and the peak energies of symmetrized EDCs ($\varepsilon_{\rm peak}$s in Fig.2a of the main paper) are also shown.  
{\bf c}, The used $\Gamma_{\rm single}(T)$ for the fitting analysis (orange circles) and the extracted $\Gamma_{\rm pair}(T)$ (green circles) are plotted. For comparison, the $\Gamma_{\rm single}(T)$ (open black squares) and $\Gamma_{\rm pair}(T)$ (open black triangles) obtained in the main paper with a different fitting scheme (Fig.5a of the main paper) are also shown. }
\label{AnotherFitting}
\label{fig1}
\end{figure*}

%\makeatletter 
%\renewcommand{\@biblabel}[1]{[S#1]} 
%\makeatother

%\bibliography{Bi2212_arc.bib}

{\bf Supplementary References}